\newcommand{\revopts}{aps,showkeys,amsmath,amssymb,nofootinbib}

\documentclass[\revopts,onecolumn,twoside]{revtex4}

\newif\ifpdf
\newif\ifdvips

\pdffalse

\dvipsfalse

\newcommand{\hropts}{,pagebackref=false}

\ifpdf

\usepackage[pdftex]{graphicx}
\usepackage{epstopdf}

\else

\usepackage[dvips]{graphicx}
\DeclareGraphicsExtensions{.eps}

\fi

\ifpdf
\usepackage[pdftex\hropts]{hyperref}
\else
\ifdvips
\usepackage[dvips,breaklinks\hropts]{hyperref}
\else
\usepackage[hypertex\hropts]{hyperref}
\fi
\fi

\bibpunct{\unskip\textsuperscript{(\hspace{-0.2em}}}{\textsuperscript{\hspace{-0.1em})}}{,}{s}{}{\textsuperscript{,}}
\makeatletter
\def\@onlinecite#1{\begingroup\let\@cite\NAT@citenum\citealp{#1}\endgroup}
\makeatother

\newcommand{\ocite}[1]{Ref.~\onlinecite{#1}}
\newcommand{\ocites}[1]{Refs.~\onlinecite{#1}}

\newcommand{\citepp}[1]{\citetext{\hspace*{0.1em}\citealp{#1}}}

\makeatletter

\newcommand{\citeaut}[1]{{\renewcommand{\hyper@linkstart}[2]{}%
\renewcommand{\hyper@linkend}{}%
\citeauthor{#1}}}

\newcommand{\citetp}[2]{%
\citeaut{#1}#2\citepp{#1}}
\makeatother

\makeatletter
\renewcommand{\p@subsection}{}
\makeatother

\makeatletter
\def\section{%
  \@startsection
    {section}%
    {1}%
    {\z@}%
    {-0.8cm \@plus -1ex \@minus -.2ex}%
    {0.5cm}%
    {\normalfont\normalsize\bfseries\raggedright\parindent\z@}%
}%
\def\subsection{%
  \@startsection
    {subsection}%
    {2}%
    {\z@}%
    {-.8cm \@plus-1ex \@minus -.2ex}%
    {.5cm}%
    {\normalfont\normalsize\bfseries\raggedright\parindent\z@}%
}%
\def\subsubsection{%
  \@startsection
    {subsubsection}%
    {3}%
    {\z@}%
    {.8cm \@plus1ex \@minus .2ex}%
    {.5cm}%
    {\normalfont\small\itshape}%
}%
\makeatother

\newcommand{\diff}{\mathrm{d}}

\newcommand{\iup}{\mathrm{i}}

\newcommand{\rtext}[1]{\hspace{0.5em}\text{#1}}

\newcommand{\equad}{\hspace{3em}}

\newcommand{\cii}{\tensor{c}}

\newcommand{\sr}{SR}

\newcommand{\slab}{$\Sigma$}

\newcommand{\scomm}{\Sigma'}
\newcommand{\scom}{$\scomm$}

\begin{document}
\title{Sagnac effect, twin paradox and space-time topology ---\\
 Time and length in rotating systems and closed Minkowski space-times}
\author{Olaf Wucknitz}

\email[e-mail: ]{olaf@astro.physik.uni-potsdam.de}
\homepage[WWW: ]{http://www.astro.physik.uni-potsdam.de/~olaf}
\thanks{}
\affiliation{Universit\"at Potsdam, Institut f\"ur Physik
 (Astroteilchenphysik), Am Neuen Palais 10, 14469 Potsdam, Germany}


\date{\today}

\begin{abstract}
We discuss the Sagnac effect in standard Minkowski coordinates and with
an alternative synchronization convention. We find that both approaches lead to
the same result without any contradictions. When applying standard
coordinates to the complete rim of the rotating disk, a 
time-lag has to be taken into account which accounts for the global
anisotropy. We propose a closed Minkowski space-time as an exact
equivalent to the rim of a disk, both in the rotating and non-rotating
case. In this way 
the Sagnac effect can be explained as being purely topological, neglecting the
radial acceleration altogether. This proves that the rim of the disk can
be treated as an inertial system.

In the same context we discuss the twin paradox and find that the standard
scenario is equivalent to an unaccelerated version in a closed space-time. The
closed topology leads to preferred frame effects which can be detected only
globally.

The relation of synchronization conventions to the measurement of lengths is
discussed in the context of Ehrenfest's paradox. This leads to a
confirmation of the classical arguments by Ehrenfest and Einstein.
\end{abstract}
\maketitle


\section{Introduction}

The interpretation of the Sagnac effect is a longstanding problem in the
theory of special relativity. Although different approaches of explanation
agree on the observable effects, which are in turn consistent with experiments,
the interpretation in the context of special relativity is still a matter of
debate, as the continuously high publication rate on the subject shows.
The problem is closely related with time measurements in moving reference
frames and especially with the synchronization of clocks at different
positions in rotating systems.

This paper is not intended to be a review of previous work on the subject but
instead tries to derive the theory and conventions from first principles in so
far as it seems appropriate in the given context. It is meant to be
more or less self-contained, so that some overlap with previous articles cannot
be avoided. We do not claim
a full axiomatic foundation of special relativity, like the one presented in
the book of \citetp{reichenbach24},\nocite{reichenbach69} of course.
For an overview of the Sagnac effect and closely related subjects, including
experiments, theoretical analysis and philosophical interpretation, we refer
to the articles of \citetp{post67}, \citetp{hasselbach93}, \citet{stedman97}
and especially the extensive discussion of synchronization issues by
\citet{anderson98}\phantom{.}
These reviews also include extended lists of references for further reading.
Very recently, \citet{rrf04} compiled a book about ``Relativity in Rotating
Frames'' containing a number of articles about the Sagnac effect, its
interpretation and related issues. The article by \citet{rizzi04} themselves
comprises another review of the subject.

The outline of this paper is as follows. First we will briefly describe
the problem the Sagnac effect poses for special relativity. After this
introduction
we will describe the principles of relativity. The Sagnac effect will be
described using the standard conventions, finding that
this 
is possible without contradictions. This proves that special relativity
can be used also on the circumference of a rotating disk. However, we will see
that the standard coordinates are not valid globally in this case, which leads
us to the discussion of a more general approach. We will derive general
coordinates and transformations which are still equivalent and compatible with
standard relativity although they use a different convention. We will
argue 
that the synchronization can be chosen arbitrarily and show that in certain
situations the choice of a non-standard convention can be more
convenient, although not necessary.

We will discuss why the standard coordinates are not valid globally in the
situation of a rotating disk, and show that the problem can be formulated
without explicitly considering the acceleration. Instead, the closed topology
of the rim of a rotating disk has to be taken into account.
Locally (which will be defined), the rim is indistinguishable from Minkowski
space. Only if our measurement process somehow encloses the complete circle,
do we notice the rotation and are able to detect any non-trivial effects. This
in a 
natural way leads us to the central point of this paper, which is the
topological 
interpretation. We will construct a (spatially) closed 
Minkowski space-time which will be revealed to be exactly equivalent to the
rim of a rotating disk. With this model at hand, we can easily avoid some
of the difficulties present in previous discussions of the Sagnac effect. This
proves that acceleration is not the main reason for the problems of
interpretation. Special relativity is valid, and the rim of the rotating disk
can even be seen as an inertial frame, as long as the radial direction is not
probed by the experiment.

After this discussion, we will turn to the twin paradox which is shown to be
closely related to the Sagnac effect, especially when it is discussed on the
rotating disk or equivalently in closed space-time.
As before, we will propose a scenario which avoids any acceleration effects
while it still leads to exactly the same physical effects as the standard
situation. In this way the twin paradox is interpreted in terms of space-time
topology as well.

Finally, we will discuss measurements of lengths, which are
related with the synchronization problem. We will find that, although the
synchronization of clocks is a matter of convention, spatial lengths of
objects at rest have a
meaning as invariant intervals in space-time only if these are measured along
lines of standard simultaneity. The measurement of lengths of moving objects,
on the other hand, is more a matter of discussion. The result will generally
depend on the details of the experiment which implicitly applies a
synchronization convention in order to define the lengths.
This will lead to our interpretation of the
Ehrenfest paradox which agrees with classical approaches but disagrees with
several recent publications on the subject.

In all our presentation we will avoid any sophisticated mathematical methods
in order not to hide the physical content behind the formalism.
We put the emphasis on a self-consistent and logical derivation of the
arguments, without hiding any implicit assumptions.
Although we tried to include references to most of the relevant publications,
we do not claim completeness in the discussion of previous work.

\section{The problem}
\label{sec:problem}

\begin{figure*}[ht]
\includegraphics[width=0.65\textwidth]{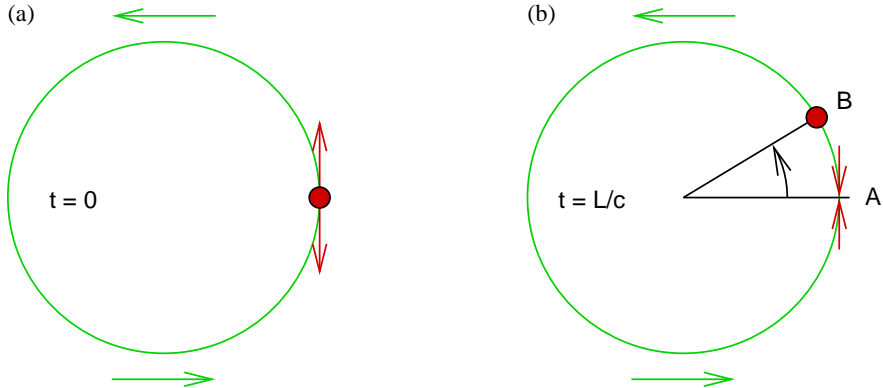}
\caption{Basic scenario of the Sagnac effect as seen from the laboratory
  frame \slab. (a) Light signals are emitted in 
  both directions by a corotating device or observer. (b) When both
  signals have completed a full round, they meet again at their starting point 
  (measured in the laboratory frame). Since the observer has moved during this
  time (from $A$ to $B$), he detected the clockwise signal some time ago but
  will detect the counter-clockwise signal only in the future.}
\label{fig:basic}
\end{figure*}

The effect in question was first proposed and knowingly measured by George
\citeaut{sagnac13a},\citepp{sagnac13a,sagnac13b,sagnac14} and was then
interpreted as the 
proof of existence of the ``luminiferous ether'' and as a measurement of
rotation relative to it.
For our discussion we assume
a simplified setup (see Fig.~\ref{fig:basic}) which differs from real
experiments in several unimportant 
aspects. See, e.g., \ocite{post67} for a discussion of the influence of more
concrete experimental setups.

On a disk of radius $R$, which can rotate
around its central axis, we arrange a collection of mirrors or wave guides
which are able to route a ray of light in a closed path (length $L=2\pi R$)
around the rim of the 
disk. For the sake of simplicity, we assume that this closed path equals the
circular rim of the disk. A device fixed to the disk is now sending
light in both directions and combines both light paths after a complete
revolution in either direction. Interference fringes are used to measure
a possible phase shift between the two light paths.
In this paper we discuss the equivalent situation of two light pulses
traveling in both directions and the measurement of their corresponding light
travel times (or their difference). The time difference, if related to the
frequency of the observed light, equals the observed phase shift of
interference fringes. 

As long as the disk is at rest with respect to the inertial laboratory
reference frame \slab, the light travel time is equal in both directions
and no phase 
shifts are observed. We now set the disk in stationary rotation (angular
velocity $\omega$, speed of the rim $v=\omega R$) and repeat the
experiment (see Fig.~\ref{fig:basic}). When viewed from the laboratory system,
the interpretation is 
simple and does not differ between non-relativistic Galilean physics and
special relativity. A time interval $\tau=L/c$ after sending, both light pulses
reach the position where they originated, because the rotating mirrors do not
change the speed of light in \slab. The corotating observer (at rest in the
rotating reference frame \scom), on the other hand, has moved by 
$v\tau$ on the circle (from $A$ to $B$ in Fig.~\ref{fig:basic}),
so that the light signal sent opposite to the rotation of the disk already
crossed 
the observer, while the signal traveling in the same direction will reach the
observer only later.
The light travel time \emph{relative to the comoving observer} (measured
either in \slab\ or \scom) does depend on the direction of light travel.
The travel times for the light moving in the same (opposite) direction
as the rotation of the disk, called co- and counterrotating from now on,
are called $\tau_+$ and $\tau_-$ in the laboratory frame \slab, and are given
by 
\begin{align}
\tau_\pm &= \frac{L}{c\mp v} \rtext{,}
\label{eq:round trip lab}
\end{align}
with a difference of
\begin{align}
\Delta\tau = \tau_+ - \tau_- 
= \frac{2vL}{c^2}\gamma^2 \rtext{.}
\label{eq:time diff lab}
\end{align}
Here we have defined $\gamma=(1-v^2/c^2)^{-1/2}$, as usually.
If $S=\pi R^2$ denotes the area of the rotating disk, this can be rewritten in
a form which is often found in the literature:
\begin{align}
\Delta\tau &= \frac{4\omega S}{c^2} \gamma^2
\label{eq:time diff omega}
\end{align}
This equation can still be used if the light path is not circular but has an
arbitrary shape with enclosed area $S$.
For this paper we prefer the form of Eq.~\eqref{eq:time diff lab}, which is
formally related to Eq.~\eqref{eq:time diff omega} by the Stokes theorem,
because we 
will later restrict ourself to the rim of the disk so that not $S$ and
$\omega$ but $L$ and $v$ will be the fundamental properties.
In the general case of non-circular light paths, Eq.~\eqref{eq:time diff lab}
can still be used as integral over the path. This form even holds
for non-rigid rotation, as will be discussed in Sec.~\ref{sec:conveyor} with a
variant of the experiment.

The trouble begins when we interpret this finding in the reference frame of the
corotating observer \scom. The distance traveled by both light signals
measured in \scom\ (in contrast to \slab) can not depend on the direction,
given the symmetry of methods to measure lengths. Both signals travel the
circumference length of the rim of the disk, measured in \scom. If the total
travel 
time, on the other hand, \emph{does} depend on the direction, it seems as if
the speed of light is not isotropic anymore which then might violate the
principles of special relativity.

The easy way out of this dilemma is denying the problem altogether, by arguing
that the comoving observer is not defining an \emph{inertial} frame so that
Lorentz transformations cannot be used, and a constant velocity of light
cannot be expected in this frame.
In the following we will show why relativity is not violated and how the
comoving frame can properly be interpreted as inertial, as long as the
experiments only probe the tangential but not the radial direction in which
the acceleration acts.
Instead of avoiding the problem, we will in this way acquire valuable insights
into the theory of special relativity, not only in the case of rotating systems
but also in general.

\section{Principles of special relativity}

The result of the experiments by Michelson and Morley is often summarized in
the form that \emph{the speed of light is the same in all inertial systems.}
The propagation of light in vacuum in any inertial system can therefore be
described by the equation
\begin{align}
\diff s^2 &=0 \rtext{,}
\end{align}
with the definition
\begin{align}
\diff s^2 &= c^2\,\diff t^2 - (\diff \xi^2 + \diff\eta^2 + \diff\zeta^2)
\label{eq:minkowski metric}
\rtext{,}
\end{align}
where $t$ is the time coordinate and $\xi,\eta,\zeta$ are Cartesian space
coordinates. Note that the principle of relativity does not postulate that
this holds in all coordinates but merely that coordinates exist in which it is
true.
This claim extends the Galilean notion of equivalence of all inertial
frames to electromagnetic phenomena.
In the following we use geometrized units in which the speed of light becomes
$c=1$ for simplicity. 

The same fact can be formulated in arbitrary generalized coordinates $\xi^\mu$,
which becomes more important in general than in special relativity.
With a metric $g$, the infinitesimal interval $\diff s^2$ can be written as
\begin{align}
\diff s^2 &= g_{\mu\nu}(\xi^\alpha)\, \diff \xi^\mu \, \diff \xi^\nu
\rtext{.}
\end{align}
Here we use the Einstein convention and sum implicitly over pairs of equal
upper and lower indices. Index $\mu=0$ stands for the time coordinate,
$\mu=1\dots3$ for the spatial ones.
In the case of standard Minkowski coordinates, the metric is Minkowskian,
$g_{\mu\nu}=\eta_{\mu\nu}=\mathrm{diag} (1,-1,-1,-1)$. Then the
coordinate $t$ measures time, and the spatial coordinates are Cartesian.
The physical theory resulting from this approach we call \emph{Special
  Relativity} or ``\sr'' in the following, regardless of the coordinates that
are 
used to describe the physical processes. This definition is important, because
often the use of certain coordinate conventions is included in the notion of
\sr, sometimes explicitly but often only implicitly. Unclear or ambiguous
definitions can then easily lead to seeming contradictions or to endless
semantic discussions.
For all our work, we assume that \sr\ is a correct theory of physical reality
in the absence of gravitation, but we will discuss different
coordinate conventions to describe the same theory.

Since the speed of light is the same in all reference frames, the interval
$\diff s^2$ must also be invariant under coordinate transformations,
especially those that correspond to a change of the reference frame.
Together with the principle of relativity, this means that Minkowski
coordinates exist for all inertial frames, and the metric has the same form 
of Eq.~\eqref{eq:minkowski metric} in all systems. Minkowski coordinates of
different inertial frames are related through the Lorentz transformations.

An important point to discuss is the physical meaning of the invariant
interval $\diff s^2$. Let us consider a clock in uniform motion. Clearly there
exists an inertial coordinate system $(t',\xi',\eta',\zeta')$ in which the
clock is at rest, i.e.\ $\xi',\eta',\zeta'=\text{const}$. In 
this system, we have $\diff s^2 = \diff t'^2$ so that $\diff s$ is a time
interval measured by this clock. Since the interval is invariant, we can as
well use any alternative coordinate system and still find that $s$ measures
the time shown by a moving clock.
An ideal clock
would be insensitive to accelerations so that it measures the integrated
$\diff s$ along its world line, the so-called proper time, regardless of
whether it is accelerated or not. This is one aspect of the principle of
locality which states that, even in accelerated systems (which are described by
coordinates different from Minkowski), for each \emph{event} or
space-time location there exists a
local comoving inertial frame. In this frame Minkowski coordinates can
be used which are therefore a good description for an infinitesimally small
region around this event. Local physics is described in this
local \emph{tangential space-time}. Note that in this theoretical
limit the principle of locality is a part of \sr, but not an
additional assumption. See the discussion in Sec.~\ref{sec:sagnac
minkowski} below for the relevance of locality in our case and for
references. 

Spatial lengths will be discussed in more detail later. Minkowski
coordinates are constructed in a way that the spatial components measure
lengths of objects \emph{at rest} directly. We will discuss this
below in Sec.~\ref{sec:length X} and generalize it in our discussion
of the Ehrenfest paradox in Sec.~\ref{sec:ehrenfest}.

\section{Sagnac effect in standard Minkowski coordinates}
\label{sec:sagnac minkowski}

Let us now return to the Sagnac effect and introduce appropriate
coordinates for the rim of the disk. We assume that the inertial laboratory
system \slab\ is given by $(t,\xi,\eta,\zeta)$, with a Minkowski metric
according to Eq.~\eqref{eq:minkowski metric}. We restrict 
ourselves to the disk plane defined by $\zeta=0$, in order to eliminate one
coordinate. We now introduce polar coordinates $r,\phi$,
measured in the laboratory frame,
\begin{align}
\xi = r \cos\phi \rtext{,} \equad
\eta = r \sin\phi \rtext{.}
\end{align}
Written in these coordinates, the metric of the same space-time is given by
the line element
\begin{align}
\diff s^2 &= \diff t^2 - \diff r^2 - r^2 \,\diff \phi^2 \rtext{.}
\label{eq:metric outside}
\end{align}
Since we are interested only in effects \emph{on the rim} of the rotating
disk, we furthermore restrict ourselves to a fixed radius $r=R$.
It is now convenient to introduce a spatial coordinate $x=r\phi$ measured
along the rim of the disk.
When restricted to the rim of the disk, the metric is now
\begin{align}
\diff s^2 &= \diff t^2 - \diff x^2 \rtext{,}
\label{eq:minkowski rim}
\end{align}
which is exactly equivalent to a 1+1 dimensional flat Minkowski
space-time. This flatness would still hold with the $\zeta$ coordinate
included. 
We learn the fundamental lesson that, as long as we are restricted to
the rim of the disk, the curvature of the circumference and the corresponding
acceleration when the disk is rotating do \emph{not} influence the physics.
Please note that this is more than just a manifestation of the principle of
locality, stating that Minkowski coordinates can be used locally.
Instead the metric is \emph{exactly} Minkowskian not only in infinitesimally
small regions but \emph{in the whole region} where the coordinates are valid.

The fact that acceleration is not relevant here was indeed mentioned before,
e.g.\ by \citet{pascual04} in a more formal mathematical way.
Real clocks and rulers would always have some small extension in the radial
direction so that the fixing of $r$ might seem very artificial.
However, it can be shown that the limit of infinitely small extension in
radial direction is well defined and equals our approach of neglecting the
coordinate altogether. This was shown in an elementary way by \citet{dieks04}
for the case of light-clocks in 
accelerated frames and in a more formal but more general way by
\citeaut{mashhoon90}.\citepp{mashhoon90,mashhoon90b,mashhoon04} As long as the measurement devices
are sufficiently small (i.e.\ much smaller than the typical scale defined by
the reciprocal acceleration), the principle of locality holds, and fixing $r$
is well justified.
A discussion including non-local effects in this sense was published by
\citetp{sorge04}. 

If space-time, which defines the physics on the rim of the disk, is equivalent
to 
Minkowski space-time, it is natural to use the standard Lorentz transformation
for the moving reference frame. Let us now set the disk into stationary
rotation with a velocity of the rim of $v$. In most what follows, we define
the rotating disk as the 
geometrical space as explained before. We do not (yet) consider a real material
disk made of solid matter. This means that we do not take into account effects
of tension and 
elasticity. Possible expansion or contraction of the disk as a result of the
motion will be discussed in the context of the Ehrenfest paradox in
Sec.~\ref{sec:ehrenfest}. 

The standard Lorentz transformation into the comoving reference frame \scom,
defined by coordinates $(t',x')$, is given by
\begin{align}
\diff t' &= \gamma \,(\diff t - v \,\diff x)
\label{eq:lotra disk 1}
 \rtext{,}
 \\
\diff x' &= \gamma \,(\diff x - v \,\diff t)
\label{eq:lotra disk 2}
 \rtext{,}
\end{align}
where, as before, $\gamma=1/\sqrt{1-v^2}$.
A fixed position on the disk with $\diff x'=0$ moves with $\diff x=v\,\diff t$,
as required.
The metric written in the new primed coordinates is the same Minkowski metric
of Eq.~\eqref{eq:minkowski rim} as in the laboratory frame, with the
consequence 
that the coordinate speed of light measured in \scom\ as $\diff x'/\diff t'$
has its universal value of $c=1$.
Now the problem of the Sagnac effect manifests itself in the following
question: 
\emph{How can this constant speed of light be consistent with the observed
different round travel times of light in different directions?}

The solution to this problem is given by the global nature of the Sagnac
effect. The round travel times do not measure \emph{local} velocities
directly, and it is not a priori clear that integrating local
effects ends up with the global behavior measured in a completely different
experiment. 
This can only be expected if the local measures (i.e.\ coordinates) do
\emph{match globally.} We will see that this is not the case for 
time on the rotating platform.

Let us return to Eq.~\eqref{eq:lotra disk 1} and its interpretation.
The occurrence of $\diff x$ as part of the expression for $\diff t'$ reflects
the fact that in \sr\ simultaneity (defined as $t=\text{const}$ \emph{or}
$t'=\text{const}$) has no frame independent meaning but differs for different
states of motion of the reference frame. This is indeed one of the central
concepts of relativity.
In the standard formulation of the theory, the shift of simultaneity between
different systems takes exactly the form of this equation. 
We will see later, that even in one \emph{given} inertial frame simultaneity is
a matter of convention and can be defined in different ways. These two facts,
that in \sr\ simultaneity depends on the state of motion and that even for a
fixed reference frame there is still freedom of conventions, should not be
confused. The former is part of the standard formulation of \sr, while the
latter, if expressed in  coordinates, leads to a more general formulation of
the 
same theory. For a fundamental discussion of these issues, we refer to the
work of
\citeaut{reichenbach24},\citepp{reichenbach24,reichenbach28}\nocite{reichenbach58,reichenbach69}
and \citet{dieks91}.

If we move along a space-like
curve with $\diff t=0$ and $\diff x=R\,\diff\phi$, we return to the same event
where we started after $\Delta\phi=2\pi$ or $\Delta x=L=2\pi R$. This is a
result of 
the periodic nature of the coordinate $\phi$ or $x$.
We can now apply Eqs.~\eqref{eq:lotra disk 1}--\eqref{eq:lotra disk 2} to
investigate 
how the primed coordinates of \scom\ change along the same integration path:
\begin{alignat}{4}
\Delta t&=\int\diff t &&= 0
&
\Delta x &=\int\diff x &&= L
\\
\Delta t' &=\int\diff t' &&= -\gamma v L
\equad
&
\Delta x' &=\int\diff x'&&= \gamma L
\label{eq:delta disk t'xi'}
\end{alignat}
We notice that, after going around the circle completely, we arrive at
\emph{different} coordinates in the primed system although we ended up at the
\emph{same event} in space-time.
The change in spatial coordinate owing to Eq.~\eqref{eq:delta disk t'xi'} is
not so surprising. We started with a periodic coordinate $x$ and could
therefore expect periodicity also in $x'$. We learn that the period is
expanded by $\gamma$, that means the circumference $L'$ measured in the moving
frame \scom\ is larger than measured in \slab,
\begin{align}
L'=\gamma L \rtext{.}
\label{eq:L'}
\end{align}
The discussion of this effect is deferred to Sec.~\ref{sec:ehrenfest}.

\begin{figure*}[ht]
\includegraphics[width=0.85\textwidth]{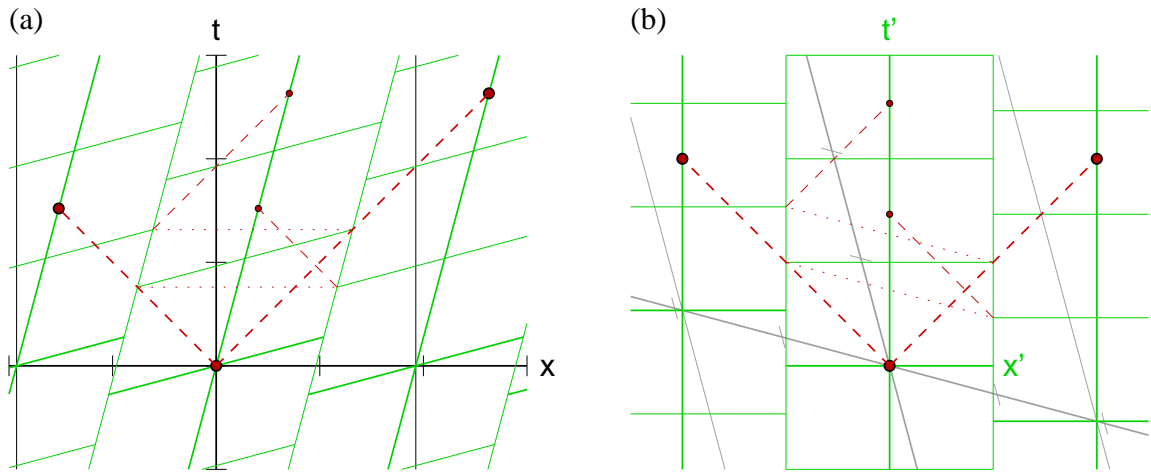}
\caption{Minkowski coordinates on the rim of a rotating disk. (a) Laboratory 
  coordinates \slab\ (upright orthogonal axes, additional thin lines at $x=\pm
  L$) and comoving coordinates \scom\ (tilted grid). (b) 
  Lorentz transformation to comoving coordinates \scom\ (upright
  orthogonal   grid) with laboratory coordinates \slab\ (tilted axes,
  additional thin lines   at $x=\pm L$). 
  The discontinuous cut was placed at comoving coordinates $x'=\pm L'/2$ with
  $L'=2\,\text{units}$. 
  The world lines of light signals starting at the origin and traveling in
  different directions are shown by dashed lines. Thick lines show the path in
  the periodic space-time, thin lines are
  folded into the fundamental coordinate interval $[-L'/2,L'/2]$.
  We notice that, although the local speed of light is $c=1$ in both
  directions, 
  the total round travel times differ for the both directions because of the
  time-lag at the cut.
}
\label{fig:spacetime sagnac}
\end{figure*}

Of fundamental importance for the Sagnac effect is the fact that the time
coordinate $t'$ changed as well, as can
be seen from Eq.~\eqref{eq:delta disk t'xi'}.
We learn that, although the primed Minkowski coordinates are valid
\emph{locally} without any 
restrictions, they can not be used as continuous \emph{global} coordinates for
the whole disk at once.
In order to use them nevertheless, we have to introduce a cut at some
arbitrary position $x'$ or $x$ (and its periodic replications).
More generally, this cut can be placed at any time-like world line
$x'(t')$ or $x(t)$. 
On the disk with this cut excluded, the coordinates are then
well-behaved and can be used to describe physics in the usual way.
If we ever want to \emph{cross} the cut in our physical experiment, we have to
take 
the jump (or \emph{time-lag}) $\Delta t'$ of the time coordinate into
account. We define $\Theta$ as
\begin{align}
\Theta=\gamma v L = v L' \rtext{,}
\label{eq:time lag}
\end{align}
so that $\Delta t'=\mp\Theta$ for co- and counterrotating paths,
respectively. If the walk around the disk does not happen instantaneously, but
takes a time interval $\tau'$ measured by a clock at rest in \scom, this
interval has to be added to $\Delta t'$.
In order to clarify the sign convention, we
repeat that if we cross the cut in positive (negative) $x'$ direction and go
around completely, the proper time interval $\tau'$ measured at the starting
position is the integrated 
coordinate time interval $\Delta t'$ \emph{plus} the time-lag $\pm\Theta$:
\begin{align}
\tau'=\Delta t'\pm \Theta
\label{eq:time lag appl}
\end{align}
In Fig.~\ref{fig:spacetime sagnac} we illustrate the coordinates including the
periodicity and discontinuity.

In the absence of any time-lag, we could have determined the round travel time
of a light signal be integrating coordinate time intervals along the
way. Because the local speed of light is $c=1$, this amounts to a round travel
time of $\tau'_0=\Delta t'=L'$, equal to the circumference length.
In reality, however, we measure the time interval at one position but the
integration (or light) paths bring us around the rim of the disk in
either direction. To 
convert the integrated local time intervals to a time interval measured at one
position, we thus have to correct for the
time-lag in the way defined by Eq.~\eqref{eq:time lag appl}:
\begin{align}
\tau'_\pm=\tau'_0\pm\Theta
\end{align}
Hence we obtain
\begin{align}
\tau'_\pm &= ( 1 \pm v)\, L'
\label{eq:tau'+-}
\end{align}
and a light travel time difference of
\begin{align}
\Delta\tau'&=2\Theta = 2vL' \rtext{.}
\label{eq:time diff'}
\end{align}
If we compare this with the results derived in the laboratory frame \slab\
from  Eqs.~\eqref{eq:round trip lab}--\eqref{eq:time diff lab}, we find
that 
\begin{align}
\tau_\pm=\gamma \, \tau'_\pm \rtext{,} \equad
\label{eq:tau'/tau}
\Delta\tau =\gamma \, \Delta\tau' \rtext{.}
\end{align}
The comoving time intervals in \scom\ are seen \emph{dilated} by $\gamma$ in
the laboratory frame \slab, just as expected from a special relativistic
discussion.
Seen from the moving frame \scom, the intervals seem to be \emph{compressed}
by $1/\gamma$. This asymmetry is not in contradiction with \sr\ but is a
direct result of the asymmetric definition of the time intervals in both
frames. In \scom\ they are proper time intervals measured by a clock at rest,
whilst the same clock is moving in \slab\ so that events at different
positions are compared. How this situation compares with the time dilation of a
clock at rest in \slab, seen by an observer in \scom, is discussed briefly in
the context of the twin paradox in Sec.~\ref{sec:time dilation} below.

The conclusion of this section is that the Sagnac effect can be explained
without any explicit physical effects of the acceleration. We confirmed that
special relativity can also be used to 
describe effects on a rotating disk. The rim of the disk can even
be seen as 
an \emph{inertial system} as long as the radial degree of freedom, in which
the acceleration acts, is not
probed by the physical experiment. An explanation in terms of acceleration can
even be excluded, because the radial acceleration cannot prefer either of the
tangential directions. When inverting the sense of rotation, the Sagnac effect
would changes its sign, although the acceleration is unchanged.

\emph{Locally}, the Sagnac effect can not be detected and the situation is
equivalent 
to a standard relativistic inertial Minkowski frame. The only difference is
that the local Minkowski coordinates cannot be extended to define a
\emph{global} 
coordinate system for the complete rim of the disk.
Note that ``locally'' here does not mean ``in an infinitely small region'' but
``in any extended region not quite surrounding the complete circle''. It is
thus rather a topological notion than one of differential geometry.
This already hints in the direction of
our interpretation of the Sagnac effect as being a purely topological
(d)effect. We will later define a physical situation which shows exactly the
same effects as the rim of a rotating disk (both locally and globally) but
which is by construction free of accelerations.

\section{Generalized Minkowski coordinates}
\label{sec:coord gen}

We found that the standard Minkowski coordinates (obtained from a Lorentz
transformation of the laboratory coordinates) do not provide a valid
continuous global system. We might therefore ask if there is an alternative
coordinate system which is better suited for the description of global effects
on the rotating disk. Generalized non-Cartesian coordinates have been
regularly used in 
pre-relativistic physics, and there is no reason not to continue the approach of
allowing non-standard coordinates in special
relativity.\citepp{dieks04} This does not mean that we have to change to
general relativity, 
because we do not discuss gravitation. In case of doubt about the correctness
of certain calculations, however, we can always escape to the \emph{formalism}
used in general relativity and define arbitrary coordinates. Physical
measurements are then defined in local
tangential space-times. When discussing the
geometry of a rotating disk in Sec.~\ref{sec:ehrenfest}, we will even
meet a curved Riemann space which is required to describe the spatial
geometry in \scom.

Before we will in the following define generalized
coordinates $(T,X)$ for a 1+1 dimensional space-time, we want to discuss our
demands for the 
properties of coordinates so that they provide a useful description of
space-time in a given inertial system.
In contrast to many other similar derivations we do
not start with a Minkowski system and determine the possible transformation to
alternative inertial systems, but instead define directly the coordinates
and metric. Transformations to and from standard Minkowski coordinates will be
discussed in the following step.

Ideally, the $T$ coordinate measures \emph{time}
and the $X$ coordinate measures \emph{length} as they are seen by an observer
at rest with respect to the inertial reference frame (e.g.\ the rotating or
non-rotating disk in our case). Without these demands, the coordinates can
only be seen as labels for 
space-time events so that the interpretation in terms of \emph{space} and
\emph{time} becomes difficult.
Our requirements on $T$ and $X$ do directly restrict the coordinate freedom by
referring to a given reference frame which is defined by
a state of motion. Any coordinate freedom left after fixing the reference
frame is then a matter of convention.

In order to understand what these notions
precisely mean, we have to define the terms very carefully.
As Ansatz for a general metric we use
\begin{align}
\diff s^2 &= A_{TT}\, \diff T^2 - A_{X X }\, \diff X^2 - 2A_{X T}\, \diff X \,
\diff T 
\label{eq:metric general}
\rtext{,} 
\end{align}
with $A_{X T}^2+ A_{X X } A_{TT} >0$ to obtain an indefinit metric, and with
$A_{TT},A_{X X }>0$ for the
convenience of simple interpretations.\footnote{We will see later that
  $A_{TT}>0$ is necessary while $A_{XX}>0$ simply avoids unnecessary
  complications without hiding any underlying problems.}
In the following we will define the properties of the metric coefficients
$A_{\mu\nu}$. In order to simplify matters, we assume that our space-time
metric is
homogeneous in $X$ and stationary (i.e.\ homogeneous in $T$), so that the
coefficients are constant.

Remember that we assume \sr\ to be correct. We do not define our metric as a
test theory of relativity but as a generalized formulation of relativity
itself.
Our approach is thus fundamentally different from the test theory of
\citetp{robertson49}, who fixed the synchronization but allowed for violations
of relativity. Later another very influential test theory was discussed by
\citeaut{mansouri77a},\citepp{mansouri77a,mansouri77b,mansouri77c} which also
included a free 
synchronization convention in addition to possible deviations from
\sr. Unfortunately, some confusion between both aspects rendered the
discussion of this theory difficult.
A test theory generalized for rotating frames is discussed by
\citetp{vargas89}.
See \citet{will92} and references therein for a
discussion of both aspects of the test theories in a number of experiments. 
No violations of special relativity have been detected so far.

\subsection{Length}
\label{sec:length X}

We assume that coordinate differences $X$ can be measured with rigid rulers
which are at rest in the reference frame and free of tensions. The state of
motion of these rulers 
does indeed \emph{define} our rest frame so that ``being at rest'' and
``having constant $X$'' are equivalent statements. By using rulers, we
assume (as a matter of fact \emph{stipulate}) that the rest length (also
called proper length) of a ruler does not change with its state of
(inertial) motion. Given the principle of relativity, this is the most
natural notion.

Absolute offsets in $X$ have no physical meaning and 
can be neglected. Offsets which are a function of $T$ would describe the
transition to a different (moving) frame.
In this way the spatial coordinate $X$ is defined uniquely for a given
reference frame.
The constancy of the metric coefficients then guarantees non-varying distances
between world lines of constant $X$, so that markings on rigid bodies do indeed
span the spatial part of the coordinate system.
Taking into account additional spatial dimensions, the situation would become
more complicated, because of possible rotations, but lead to the same
conclusions.

Note that, for the moment, we do not define a physical process to measure
lengths. We use rigid rulers (which we assume to exist) both to define the
rest frame and to measure lengths.

\subsection{Time}

Time intervals $\Delta T$ at constant $X$ can be measured as proper time
intervals of a clock at rest at $X$. We therefore demand $\diff s=\diff T$
for clocks with $\diff X=0$, which with Eq.~\eqref{eq:metric general}
leads directly to 
\begin{align}
A_{TT} = 1 \rtext{.}
\end{align}
Global offsets of the coordinate $T$ are again not
relevant. In contrast to the $X$ coordinate, where differential offsets as a
function of $T$ are fixed by the choice of a given rest frame, differential
offsets of $T$ as a function of $X$ can \emph{not} be defined unambiguously
without additional assumptions.
We can put a collection of clocks at different positions, so that each clock
measures $T$ at a certain $X$, but the relative synchronization of clocks at
different positions is a matter of convention.
In non-relativistic physics, the absolute nature of simultaneity automatically
fixes this synchronization, which is then even independent of the reference 
frame. This reflects the fact that space and time are separate entities in
Galilean physics. In \sr\ we \emph{define} simultaneous events by $\Delta
T=0$. Simultaneity therefore inherits its conventional nature from the
synchronization.

\subsection{Definitions of the velocity of light}

The foundation of special relativity is the fact that the speed of light is
equal in all inertial reference frames. In order to measure, or even define,
the \emph{one-way} speed of light as a coordinate velocity, we have to
adopt some convention for the 
synchronization of clocks at different places.
Assuming that our coordinate system reflects this convention of time
synchronization, we can now define the one-way velocities of light in the
positive and negative $X$ direction as
\begin{align}
c_+ &= +\left. \frac{\diff X}{\diff T}\right|_{\diff X>0} \rtext{,}
\label{eq:c+}
 \\
c_- &= -\left. \frac{\diff X}{\diff T}\right|_{\diff X<0} \rtext{,}
\label{eq:c-}
\end{align}
where both derivatives are taken along the world line of a light signal, which
has $\diff s^2=0$.
It is clear that if we add to $T$ an offset which is a function of $X$, this
will generally change $c_+$ and $c_-$. Therefore the one-way velocity of light
is in no way uniquely defined, neither in theory nor in experiment. The
situation only changes once we do agree on a 
synchronization convention, i.e.\ a time coordinate $T$ and the corresponding
metric coefficients.
This point is not recognized in all
publications on the subject but particularly stressed in the discussion of
\citet{vetharaniam93} and \citet{anderson98}\phantom{.}

The \emph{two-way} velocity of light $\cii$, on the other hand, is defined by
sending a light signal to 
\emph{and fro} (e.g.\ using a mirror at fixed distance) and dividing the total
traveled distance by the total time \emph{measured at one position.} In this
way we rely on the time measurement only at one fixed position and can avoid
the synchronization problem. The resulting quantity is therefore uniquely
defined and does not depend on conventions.
The length is the same in both directions while the travel times add up, so
that we find
\begin{align}
\frac{2}{\cii} &= \frac{1}{c_+} + \frac{1}{c_-} \rtext{.}
\label{eq:c2}
\end{align}
The condition for a null line can be written by setting the line element in
Eq.~\eqref{eq:metric general} to zero. After dividing by $(\diff X)^2$ 
we obtain the quadratic equation
\begin{align}
A_{TT} \left(\frac{\diff T}{\diff X}\right)^2 - 2A_{XT}\, \frac{\diff T}{\diff
  X} - A_{XX} &= 0 \rtext{,}
\end{align}
with the solutions
\begin{align}
\left.\frac{\diff T}{\diff X}\right|_\pm &= \frac{A_{X T}\pm\sqrt{A_{X
    T}^2+A_{X X } A_{TT}}}{A_{TT}} \rtext{,}
\label{eq:1/c+-}
\end{align}
one of which is positive and one negative. These correspond to light travel in
positive and negative $X$ direction and thus to $\pm1/c_\pm$.
With Eqs.~\eqref{eq:c+}--\eqref{eq:c2},
the two-way speed of light becomes
\begin{align}
\cii &= \frac{A_{TT}}{\sqrt{A_{X T}^2+A_{X X } A_{TT}}} \rtext{.}
\label{eq:c2b}
\end{align}

\subsection{Constancy of speed of light}

As discussed above, only the two-way speed of light has a convention
independent meaning. It is indeed this $\cii$ which was measured in the
Michelson-Morley experiment and found to be constant, independent of direction
(we discuss only one spatial coordinate so that this does not apply) and
independent of the motion of the reference frame in which it is measured.
This extends the principle of relativity
to electromagnetic phenomena. All inertial reference frames are
equivalent and cannot be distinguished even when using light to probe the
space-time. 

Since we have $\cii=1$ measured in one inertial frame, the same equation has
to hold in any other frame as well. In the notation of this paper, we have
even \emph{defined} our units in a way to obtain $\cii=1$. This is actually
very 
similar to the current SI definition of the meter as the length which light
travels in a certain time. In this sense the theory of special relativity
can be derived from two principles: (a) equivalence of all inertial frames and
(b) the finiteness of the speed of light. If the speed of light were
infinite, the Lorentz transformation would degenerate to the Galilei
transformation. The principle of relativity would still be satisfied then.

With Eq.~\eqref{eq:c2b}, the condition $\cii=1$ leads directly to
\begin{align}
A_{X X } A_{TT} + A_{X T}^2 &= A_{TT} ^2 \rtext{.}
\end{align}
Recall that we have not analyzed the measurement of lengths in
detail. We simply assumed it can be done with rigid rulers in their
rest frame.
We then used the Michelson-Morley result of the constancy of the
speed of light. This does actually not describe the propagation of light alone
but the relation of the light propagation to the length of rigid rulers.
In this paper we leave the question open whether this tells us anything at all
about the propagation of light or if it is just a result of the fact that
rigid rulers are held together by electromagnetic forces (i.e.\ by light) so
that they naturally have to Lorentz contract when set into motion in a way
which keeps the measured speed of light constant.

In any case we can now use two-way light travel times to measure lengths
(``light-ruler'') and, vice versa, the periodic reflection of light between
mirrors of known distance to measure time (``light-clock''). We can choose
whichever measurement procedure can be analyzed most easily, because the
principle of relativity ensures that they all lead to the same result.
This approach will be used later in the discussion of the Ehrenfest paradox in
Sec.~\ref{sec:ehrenfest}.

\subsection{Coordinates and transformations}

The conditions that $T$ and $X$ measure time and length, respectively, plus
the principle of relativity including the results
of the Michelson-Morley experiment provide us with the conditions $A_{TT}=1$
and $A_{X T}^2+A_{X X }=1$. We therefore have only one free
parameter $A=A_{XT}$ with $|A|<1$ left and can write the metric of
Eq.~\eqref{eq:metric general} as 
\begin{align}
\begin{aligned}
\diff s^2 &= \diff T^2 - (1-A^2)\, \diff X^2 - 2A\, \diff X\, \diff T
 \\
&= (\diff T-A\, \diff X )^2 -\diff X^2 \rtext{.}
\end{aligned}
\label{eq:metric general 1}
\end{align}
The one-way speed of light can now be derived from
Eqs.~\eqref{eq:c+}--\eqref{eq:c-} with Eq.~\eqref{eq:1/c+-} and
takes the particularly simple form
\begin{align}
c_\pm &= \frac{1}{1\pm A} \rtext{.}
\label{eq:c+-b}
\end{align}
For $A=0$, our general coordinates are equivalent to standard Minkowskian ones.
Soon we will learn that the parameter $A$ describes the synchronization
convention. 

The generalized Lorentz transformation with respect to a Minkowski rest frame
$(t,x)$ can now be derived from two conditions: 
(a) The proper time interval $\diff s^2$ is invariant, (b) a comoving point
with 
$\diff X =0$ travels with a velocity $v=\diff x/\diff t$ in the laboratory
frame. This 
leads after some simple algebra to the following transformation and its
inversion.
\begin{alignat}{2}
\diff T &= \gamma \left[ (1-Av)\, \diff t - (v-A)\, \diff x \right]
\label{eq:lorentz gen first}
&\equad
\diff t &= \gamma \left[ \diff T +(v-A)\, \diff X \right] 
\\
\diff X &= \gamma \left[ \diff x - v\,\diff t \right]
&
\diff x &= \gamma \left[ (1-Av) \,\diff X + v\, \diff T \right]
\label{eq:lorentz gen last}
\end{alignat}
This transformation could be written even more generally as transformation
between two systems which \emph{both} have their own arbitrary synchronization
parameters $A$ and $A'$.

The relation to standard Minkowski coordinates in the same frame can be
derived from Eqs.~\eqref{eq:lorentz gen first}--\eqref{eq:lorentz gen last}
with $v=0$ and $\gamma=1$:
\begin{align}
\diff T = \diff t' + A \,\diff x'  \equad
\diff X = \diff x'
\end{align}
The alternative coordinates $(T,X)$ differ from the standard Minkowskian ones
$(t',x')$ only in a 
position-dependent offset of time, i.e.\ in the synchronization convention of
clocks. This justifies that we call $A$ the ``synchronization parameter''.

Without additional assumptions, the parameter $A$ can be
chosen freely without coming in conflict with experiments. However, not all
synchronization conventions will lead to simple physical laws written in
those coordinates. Some researchers dispute this notion, often by adding
implicit and sometimes obscure assumptions, which are compatible only with
specific synchronization schemes.
Note that the question of whether the
synchronization is conventional or not is itself a matter of convention and
depends on the demands we put on a coordinate system in order to find it
acceptable. The point is that if we want to find the ``true''
synchronization parameter, we have to define what this means. Different
definitions will naturally lead to different conventions.

\citet{einstein05} himself argued that the time at $A$ and $B$ can not be
compared directly, but that a general ``time'' can be \emph{defined} by
stipulating that the light travel time from $A$ to $B$ is the same as that
from $B$ to $A$. He was therefore well aware of the fact that this is just a
convention, although he did not discuss alternative definitions.
For our presentation we want to avoid any unnecessary assumptions and treat
the synchronization as conventional.
In some sense not even our requirements that $X$ measures length and $T$
measures time are truly
necessary. They do, however, have a direct physical meaning in that they allow
a simple interpretation of the coordinates. This is fundamentally different
from the remaining coordinate freedom parameterized by $A$.

The synchronization can also be treated in terms of gauge theory. 
\citet{minguzzi02} presented a very clear and simple discussion and
additionally\citepp{minguzzi03} a far more general and formal treatment.
Further thoughts about synchronization in more general systems, including
gravitation and acceleration, are presented by \citeaut{goy96b} for the
rotating disk,\citepp{goy96b,goy96c} for clocks on earth,\citepp{goy97} and for
clocks orbiting in general relativity.\citepp{goy97d}

In the following we will discuss the standard synchronization, which is
especially well suited for any local effects, and an alternative
convention, which simplifies the discussion on the rim of a rotating disk.
Both these conventions are in some publications described as \emph{the only
  possible} synchronization schemes, because different criteria for the
(non-)acceptance are used.
Within \sr\ experiments cannot be used to select the \emph{true}
synchronization. They can only compare synchronization performed with
different methods (see the discussion in the following section).

\subsection{Standard isotropic clock synchronization}
\label{sec:standard sync}

The standard method for the synchronization of clocks is the so called
Einstein synchronization. To synchronize clock 1 with 2 (which for our purpose
are assumed to be
at rest relative to each other), we send a light signal from clock 1 (sent at
$t_A$) to clock 2 (received at $t_B$) \emph{and 
  back} (received by clock 1 at $t_C$) and set the clocks in a way that
$t_B=(t_A+t_C)/2$, so that the 
light travel time from 1 to 2 (measured as the difference in coordinate times
at 1 and 2) is the same as the time from 2 to 1. Using this \emph{convention,}
the speed of light necessarily becomes isotropic because this very assumption
was used in the definition.
In terms of our generalized coordinates, this leads to $A=0$, as
Eq.~\eqref{eq:c+-b} shows.
In the following, we want to refer to the synchronization
convention $A=0$ as \emph{standard, symmetric} or \emph{isotropic}
  synchronization, no 
matter how this synchronization is actually achieved in an
experiment.

An alternative method, which leads to the same convention, is the ``slow
transport of clocks''. We take a third clock, bring it 
to clock 1 and synchronize the two, which is trivial when they are located at
the same position. This third clock is then moved
\emph{slowly} to clock 2. Clocks 1 and 2 are synchronized if the moving clock
shows the same time as clock 2 when it arrives there. The limit of infinitely
slow transport exists and provides a well defined synchronization procedure as
we will show now.

The proper time interval $\Delta s$ of the moving clock can be calculated from
the metric of Eq.~\eqref{eq:metric general 1}:
\begin{align}
\Delta s^2 &= \Delta T^2 - (1-A^2) \,\Delta X^2 - 2A \,\Delta X \, \Delta T
\end{align}
On the way, we have $\Delta X=w\,\Delta T$ with $w$ denoting the coordinate
velocity of the moving clock. After moving the comparison clock to clock 2, we
compare the two. Clocks 1 and 2 show coordinate time $T$, so that the
difference between the interval measured by clocks 1+2 ($\Delta T$) on one side
and the transported clock ($\Delta s$) on the other side is
\begin{align}
\Delta T - \Delta s
 &= \frac{1-\sqrt{1-(1-A^2)\,w^2-2Aw}}{w}\, \Delta X
\rtext{.}
\end{align}
The limit of small $w$ exists as a constant difference:
\begin{align}
\lim_{v\to 0}\, (\Delta T - \Delta s) &= A\, \Delta X
\end{align}
The clocks are called synchronized if this difference vanishes, which is
equivalent to $A=0$. We
see that slow clock transport in either direction is equivalent to Einstein
synchronization.

Further experimental methods can be used to perform the clock synchronization
according to the same convention. These have in some cases be misinterpreted as
determinations of the ``objective'' synchronization.  A simple mechanical
method will at some 
position send two particles of equal rest mass in opposite directions with
velocities chosen in a way to obtain a total momentum of zero. This can
unambiguously be accomplished by starting with the two particles at rest (zero
momentum) and then use some internal process of the two-particle system to
separate and accelerate them with the total momentum conserved, e.g.\ with
the release of a taut spring. The
particles have some 
internal clocks and after the same proper time interval, they set the clocks
which they pass at this moment to the same time.
This scenario can be analyzed most easily in terms of the corresponding
velocity four-vectors because of their close relation to the momentum. These
vectors can be defined as derivatives of arbitrary 
coordinates with respect to the proper time of the moving particles:
\begin{align}
u^\mu &= \frac{\diff x^\mu}{\diff\tau}
\end{align}
In terms of the coordinate velocity $w=\diff X/\diff T$, we can derive the
four-velocity components using the metric of Eq.~\eqref{eq:metric general 1}
and the fact that $\diff s=\diff\tau$:
\begin{align}
u^T = \frac{1}{\sqrt{1-(1-A^2)w^2-2Aw}} \equad u^X = \frac{w}{\sqrt{1-(1-A^2)w^2-2Aw}}
\label{eq:four-vel vs coord-vel}
\end{align}
The two separating particles would, by definition, have equal $u^T=\diff
 T/\diff\tau$ and, due to conservation of the total zero momentum,
 opposite $u^X$. 
With $w=u^X/u^T$, this leads directly to opposite coordinate velocities for the
two particles and therefore, following Eq.~\eqref{eq:four-vel vs coord-vel} for
equal $u^T$, to $A=0$.

Note that the momentum conservation can be discussed in a coordinate
independent (covariant) way by using projections.\footnote{These projection
  will again be used when discussing lengths in Sec.~\ref{sec:lengths
    rest}.} 
If we denote the four-velocity of an observer at rest as $p^\mu$ (in our
coordinates $p^T=1$, $p^X=0$), we can write 
the time and space projections of $u^\alpha$ ($\bar u^\alpha$ and $\tilde
u^\alpha$, respectively) as 
\begin{align}
\bar u^\alpha = p^\alpha\, (p^\mu u_\mu)\rtext{,} \equad
\tilde u^\alpha = u^\alpha - \bar u^\alpha\rtext{.}
\label{eq:proj}
\end{align}
The time projection of the four-momentum measures the energy, the space
projection the spatial momentum. The two particles must have equal $\bar
u^\alpha$ but opposite $\tilde u^\alpha$.

The fact that this synchronization method leads to the symmetric convention
does not come by surprise, given the fact that the four-velocity vectors live
in the local tangential space-time. If the symmetry of the two vectors
is inherited by the coordinates, this must necessarily
lead to symmetric synchronization.

\begin{figure*}[ht]
\includegraphics[width=0.9\textwidth]{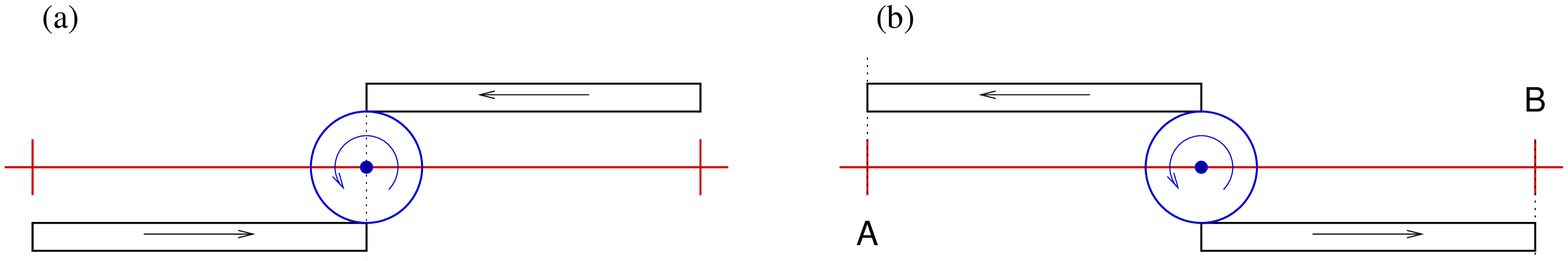}
\caption{The synchronization method of \protect\citetp{spavieri86}. (a) Two
  rods start 
  to move with the same velocity, guided by contact to a rotating
  wheel. (b) When the ends of the rods reach the clocks at $A$ and $B$,
  which are at equal distance from the central wheel, they are synchronized to
  the same time. In \sr\ this technique is equivalent to Einstein
  synchronization as shown in the text.}
\label{fig:mechsync}
\end{figure*}

A variant of this method was discussed in the work of \citetp{spavieri86}. He
did not consider two symmetrically separating particles but instead two rods
moving in opposite directions, whose equal velocities are prescribed by a
rotating wheel which is in contact with both rods (see
Fig.~\ref{fig:mechsync}). The two rods start moving 
at the same time at the same position, and when their ends reach markings which
are at both sides at equal distances, the clocks at these markings are set to
the same time. The author argued that the rods have the same velocity, so that
this method unambiguously fixes the synchronization.
This argument does not actually hold, however, if one takes into account that
only the spatial components of the four-velocities are equal but not
necessarily the coordinate velocities $\diff X/\diff T$. Assuming equality of
the latter would take symmetric synchronization for granted and thus lead to a
circular argument. Nevertheless, this setup provides a natural method to
perform the isotropic synchronization in an experiment.

We found that there are many methods to perform the standard symmetric
synchronization of clocks. This convention is indeed preferred in that it
leads to isotropic coordinates. No other \emph{locally} realizable convention,
which does not refer to some external system (like the laboratory in the
rotating disk case), can 
play this role as preferred convention. In terms of our parameter $A$ this can
be formulated as follows. If we
assume that there is one ``special'' convention, this can only be $A=0$,
because for any other value $A\ne0$ we could as well have taken $-A$. If all
frames are equivalent, only the isotropic convention is preferred before all
others.
We will see later that special global properties of space-time may lead to
the selection of preferred frames so that other conventions might be seen as
more convenient in such a case.

Assuming validity of \sr, several methods
lead to the same isotropic synchronization and thus to the standard Minkowski
metric. The equivalence of these different \emph{methods}, which in \sr\ lead
to the 
same \emph{convention,} can be tested in experiments. It should be kept in mind
that 
these are then tests of \sr\ but not experiments to determine the ``correct''
convention. See, e.g., the discussion by \citet{croca99} for 
an example where the measurement actually compares slow clock transport with
Einstein synchronization. Even the historical measurement of the speed of
light by Ole R{\o}mer, using the retardation of the observed orbits of the
satellites of Jupiter, can not be interpreted as an unambiguous measurement of
the one-way speed of light. It really measures this speed by comparing two
slowly  transported clocks, represented by the orbits of Jupiter's satellites
and the laboratory clocks on earth.

\subsection{Global synchronization}
\label{sec:global sync}

Besides the symmetric convention leading to Minkowski coordinates, there is
\emph{in our scenario} another special case of the
synchronization parameter. If we choose $A=v$, the generalized Lorentz
transformation of Eqs.~\eqref{eq:lorentz gen first}--\eqref{eq:lorentz gen
last} takes the following form: 
\begin{alignat}{2}
\label{eq:lorentz global sync first}
\diff T &= \gamma^{-1}\, \diff t &
\diff t &= \gamma \,\diff T \\
\diff X &= \gamma \,( \diff x - v\,\diff t ) & \equad
\diff x &= \gamma^{-1}\, \diff X + v\gamma\, \diff T
\label{eq:lorentz global sync last}
\end{alignat}
In the Sagnac experiment this special convention avoids the
discontinuity in the time coordinate so that the coordinates are valid
\emph{globally} in the rotating frame \scom. Simultaneity with respect to the
new coordinates is equivalent to standard simultaneity in the laboratory frame
\slab.
The periodicity of $X$ with $L'$ is unchanged in this synchronization.

Note that we can define this special convention not only by referring to the
external laboratory frame but also purely internal to the rim of the rotating
disk. This special value of the parameter $A$ is the only one which avoids
time-lags and thus allows the definition of a \emph{global} time
coordinate. In order to achieve this, it is not sufficient to use local
measurements\footnote{Again not meaning infinitesimally small regions but
  extended regions not surrounding the circle.} but we must instead
consider world lines which surround the circle completely. Concrete
experimental scenarios which can be used to perform this global
synchronization shall not be discussed here.
Locally, this
synchronization convention is in no way preferred, in agreement with the
arguments presented above.

\begin{figure*}[ht]
\includegraphics[width=0.75\textwidth]{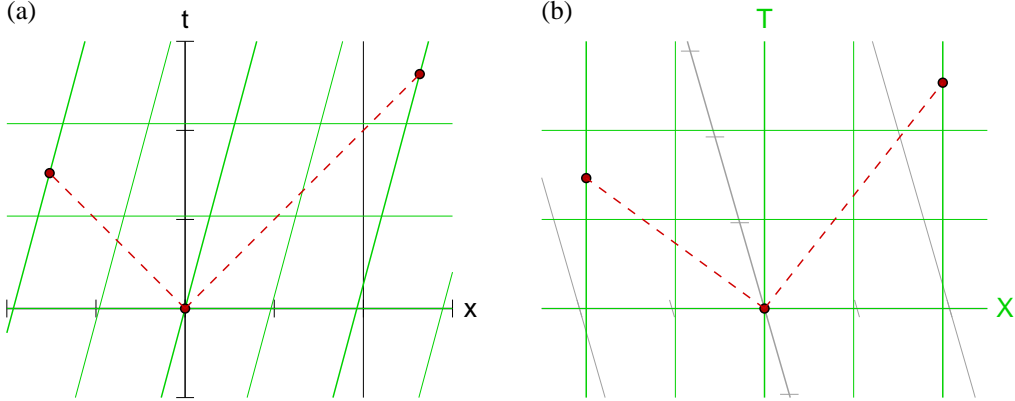}
\caption{Alternative coordinates on the rim of the rotating disk. (a)
  Laboratory coordinates \slab\ (upright orthogonal axes, additional
  thin  lines at $x=\pm L$) and  
  comoving coordinates \scom\ (tilted grid, thick lines at $X=nL$). (b)
  Generalized Lorentz transformation to comoving coordinates \scom\ (upright
  orthogonal grid) with 
  laboratory coordinates \slab\ (tilted axes, additional thin lines at $x=\pm
  L$). 
  Both coordinate systems are valid globally without discontinuities.
  The world lines of light signals starting at the origin and traveling in
  different directions are shown by dashed lines.
  In \slab\ $(t,x)$ the coordinate speed of light is isotropic, in \scom\
  $(T,X)$ it is   anisotropic. 
}
\label{fig:spacetime sagnac2}
\end{figure*}

The price we pay for the advantage of a global coordinate system is high; the
coordinates are not symmetric, and the local coordinate velocity of light is
not isotropic anymore. 
We can calculate $c_\pm$ either directly from Eq.~\eqref{eq:c+-b} or as a
special case of Eq.~\eqref{eq:four-vel vs coord-vel} for diverging
four-velocity $(u^T,u^X)$.  Both lead to  
\begin{align}
c_\pm = \frac{1}{1\pm v} \rtext{.}
\end{align}
On the other hand, the global coordinates make the interpretation of the
Sagnac effect much simpler (see Fig.~\ref{fig:spacetime sagnac2} for an
illustration). A globally preferred reference frame does exist, so 
that it is no surprise that the coordinates lose some of their symmetries when
we are in a different (i.e.\ rotating) system. In this way the global
asymmetry is formulated as a local asymmetry of coordinates.
In reference frames moving with respect to the preferred frame, the coordinate
speed of light is anisotropic, which leads to the different round travel times
on the rotating disk, $\tau'_\pm = (1\pm v) L'$.
This is exactly equivalent to the former result of Eq.~\eqref{eq:tau'+-} which
confirms our view that the synchronization is merely a convention and that the
Sagnac effect can be explained regardless of the synchronization used.

\section{Topology}
\label{sec:topology}

In the previous sections we showed that locally (when not surrounding the 
disk completely), the rim of the rotating disk is equivalent to a uniformly
moving inertial system in 1+1 dimensions. No frame is preferred and standard
Minkowski coordinates can be used without restrictions.
Only if we allow world-lines to go around completely, are we able to detect
the rotation of the disk. We find that there is a \emph{globally} preferred
reference frame, given by the system in which the disk is not rotating.
This frame is preferred only as a result of the boundary conditions which are
in turn an imprint of the closed topology.

It is instructive to investigate how the synchronization of clocks depends on
the path along which it is performed. To simplify matters, we discuss the slow
transport of clocks, which we know is equivalent to any other isotropic
synchronization technique like the Einstein synchronization.
Especially we want to concentrate on closed synchronization paths. In a
standard open Minkowski space-time, we know that a clock transported slowly
along \emph{any} closed path beginning and ending at a fixed second clock
will always stay synchronized with this latter reference clock.
This implies that for open paths the 
result does not depend on the path followed, so that a unique and global
synchronization of clocks at all positions can be achieved in this way.

In general accelerated systems like the complete rotating disk (including the
radial direction), the
transported clock will generally show deviations from the fixed reference
clock. The measured time differences can generally have any value and will
depend on the path taken. Changing the path continuously will change the
measured time difference continuously as well. In such a situation it is quite
natural to interpret this desynchronization as a result of the accelerations
in the (non-inertial) reference frame. Even if we restrict ourselves to small
but 
finite regions, do the clocks desynchronize. This is the reason why such
systems can \emph{at large} not be described as inertial systems in the
context of \sr. 

The situation is very different on the rim of the rotating disk, even if we
extend the discussion to the rotating cylinder by including the axial
coordinate $\zeta$ which we have neglected so far. 
The time difference $\Delta t'$, measured after moving one of the
clocks around,
will alway be zero as long as the path does not surround the disk (or
cylinder) completely.
For general paths, which can wind around arbitrarily often, we find that the
time difference can take only a discrete set of values, which are all multiples
of $\Theta=v L'$ as calculated in Eq.~\eqref{eq:time lag}.
The time difference (reading of the traveled clock minus reading of the fixed
clock) is
\begin{align}
\Delta t' = -n \Theta \rtext{,}
\end{align}
where the number $n\in\mathbb{Z}$ describes the total number of windings in the
positive $X$ direction.
The situation is now fundamentally different from the general case discussed
above. As long as we do not allow surrounding the rim completely, we are not
able to detect the rotation and are thus allowed to describe the reference
frame as an inertial system.
The time difference only depends on the winding number $n$ which can be
interpreted in the 
context of topology as a label for the homotopy class of the clock's path in
the multiply connected space-time defined by the periodic coordinate $X$.

This provides us with the recipe to leave the concrete scenario of the
rim of the rotating disk, where some readers may still not be convinced that
the acceleration has no direct physical effect, and move to a more general 
scenario where it is clear from the beginning that in extended regions the
space-time is flat and can be described by 
Minkowski coordinates.
Said in
physical words, we will construct inertial systems 
which have the same (local and global) properties as the space-time defined by
the rim of the rotating disk.

\begin{figure*}[ht]
\includegraphics[width=0.6\textwidth]{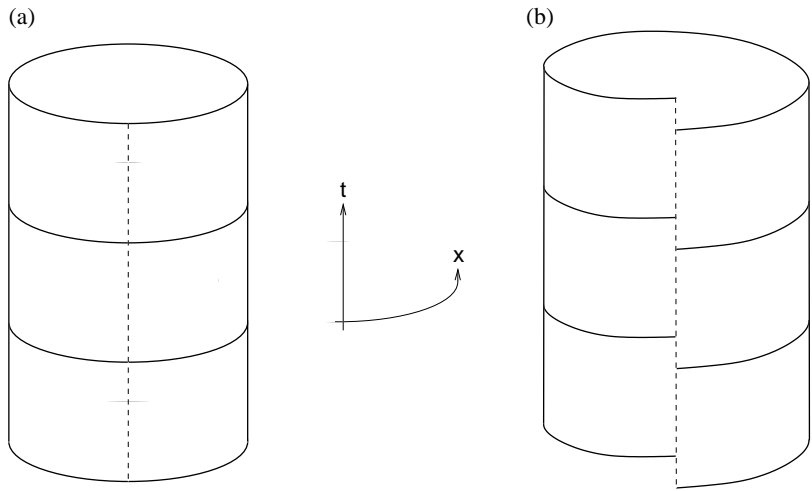}
\caption{Geometry of closed 1+1 dimensional Minkowski space-time.
Time $t$ is measured vertically, space coordinate $x$ around the cylinder.
(a) Identification of $(t,0)$ with $(t,L)$, glued together without time
jump. This corresponds to a non-rotating disk \slab. (b) Identification of
$(t',0)$ with $(t'-\Theta,L')$. 
This version corresponds to the rotating disk in the comoving frame \scom.
The elliptical curves/spirals are lines of constant $t$ and $t'$,
respectively.
Note that in both cases the cylinder is embedded in a higher dimensional
space-time only to illustrate the cylindrical topology. There is no
intrinsic curvature.
}
\label{fig:closed mink}
\end{figure*}

Let us start with a
standard Minkowski frame \slab\ in 1+1 ($t$ and $x$) dimensions and compactify
this 
space-time by making it periodic. We identify $(t,0)$ with $(t,L)$, or more
generally $(t,x)$ with $(t,x+L)$, in order to obtain
a space-time which is $L$-periodic in $x$ but otherwise still Minkowskian.
See Fig.~\ref{fig:closed mink}\,(a) for an illustration in which the space-time
is shown as a cylinder embedded in 2+1 space-time. The cylindrical shape is
used only for illustration purposes. The space-time itself is by construction
intrinsically flat.

We find that this space-time has the same topology and metric as the one
on the rim of a disk at rest which was discussed before. Exactly the same
effects can therefore be shown to exist in our periodic space-time, using the
same 
formalism as above. A moving inertial system \scom\ can again not be described
globally by standard Minkowski coordinates because global isotropic clock
synchronization is not possible.
The spatial extension (or circumference) of the space-time in the moving frame
is $L'=\gamma L$ as before (see also Sec.~\ref{sec:ehrenfest}) and becomes
minimal for $v=0$. 
Round travel times for light signals going in different directions will (for
$v\ne0$) be different as on the rotating disk, so that a Sagnac effect will
again be observed. Contrary to the scenario on the rotating disk, there is no
chance to blame acceleration for the observed effects, because the reference
frame is inertial by construction, with no acceleration at all.
This confirms and justifies our notion of inertial frames
on the rim of a rotating disk.

Instead of identifying $(t,0)$ with $(t,L)$, we could also have included an
additional time-lag and build an alternative geometry by identifying
$(t',0)$ with $(t'-\Theta,L')$, as shown in Fig.~\ref{fig:closed
  mink}\,(b). In this way we utilize the freedom the closed topology provides
us with. Note that the time step is restricted to values smaller than the space
periodicity, i.e.\ $|\Theta|<L'$, in order to avoid time-like closed curves
and to 
preserve causality and allow a free will for beings living in this
space-time. 
With the same argument we reject the possibility of an identification of
different times at the same position.
What would the effects in a geometry with time-lag be?
We can compare with Eq.~\eqref{eq:delta disk t'xi'} to
find that the situation in primed coordinates is the same as before on the
\emph{rotating} disk in comoving Minkowski coordinates in the frame
\scom. Remember also 
Fig.~\ref{fig:spacetime sagnac}\,(b) for an illustration and compare with
Fig.~\ref{fig:closed mink}\,(b).
Adding a time-lag in $t'$ therefore seems to set the global system into
uniform motion.
We can now apply the inverse of the Lorentz transformation of
Eqs.~\eqref{eq:lotra disk 1}--\eqref{eq:lotra disk 2} in order to arrive at the 
frame without time-lag and with global Minkowski coordinates.
This confirms that the time-lag in the compactified Minkowski
space-time is a measure for its global ``rotation'', equivalent to the
geometry on 
the rim of the rotating disk.
The two approaches of Fig.~\ref{fig:closed mink}\,(a) and (b)
are therefore really equivalent and differ only in respect to the reference
frame in which the compactification is performed.

We
could now repeat the whole discussion of the Sagnac effect presented above. 
We would again find travel time differences (Secs.~\ref{sec:problem} and 
\ref{sec:sagnac minkowski}) and would rediscover the possibility of using
alternative coordinates with a global synchronization of clocks
(Sec.~\ref{sec:global sync}).
We see that the situation in a closed space-time is significantly different
from the scenario normally 
discussed in special relativity with an open topology where no frame is
preferred.
When applied to a spatially closed space-time,
the standard relativistic formulation is restricted to be valid only locally
but not globally. When going around completely, discontinuities in standard
coordinates are to be expected and must be taken into account. Alternatively,
the global clock synchronization can be used, but physical laws will look
different when written in these anisotropic coordinates.

Additional aspects of the closed space-time will be discussed later in the
context of the twin paradox (Sec.~\ref{sec:twins closed}) and the Ehrenfest
paradox (Sec.~\ref{sec:ehrenfest}).

\section{Fiber optic conveyor experiments}
\label{sec:conveyor}

\begin{figure*}[ht]
\includegraphics[width=0.7\textwidth]{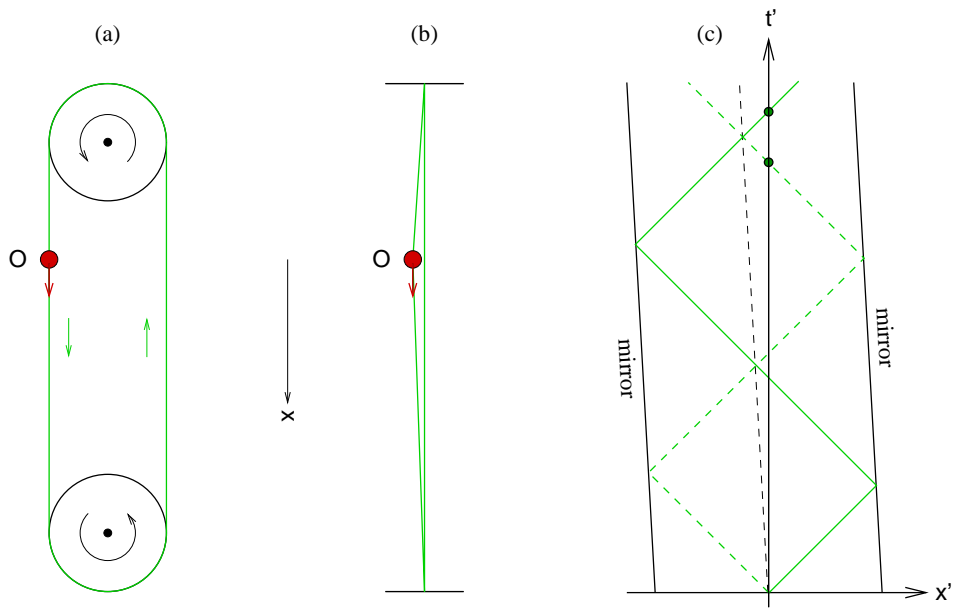}
\caption{The fiber optic conveyor Sagnac experiment. (a) Original setup. An
  optical fiber moves around two wheels. The observer $O$ moves with the fiber
  along a straight line.
  (b) Equivalent arrangement where the light guidance by the fiber around
  the wheels is replace by the reflection on two mirrors. (c)
  Space-time diagram in the comoving inertial frame \scom. The light signals
  (solid 
  and dashed zigzag lines) are reflected by the moving mirrors. Along the
  straight dashed line ($x=\text{const}$) the round travel times would be
  equal. The observer has instead $x'=\text{const}$, where the travel times
  differ in the same way as in the original Sagnac scenario.}
\label{fig:conveyor}
\end{figure*}

A modified Sagnac experiment was carried out by \citet{wang03}\phantom{.} We
illustrate the fundamental arrangement in 
Fig.~\ref{fig:conveyor}\,(a). The light is guided by an optical fiber which 
runs around two wheels and moves with a certain velocity $v$. The observer is
comoving with a straight part of the fiber and sends light signals in both
directions.
Different geometries with different enclosed areas etc.\ are tested in
\ocite{wang03} with the result that the observed time difference (to first order
in $v$) is always
$\Delta t=2vL$ for a fiber of length $L$, independent of the specific
geometry and whether parts of the fiber are moving uniformly or
circularly. This hints in the same direction as our interpretation. The time
difference is not a result of acceleration (or of enclosed area) but only of
velocity and length, which in this scenario is discussed in some detail by
\citetp{tartaglia04}, who come to conclusions quite similar to ours, namely that
acceleration is not the prime reason for the Sagnac effect, but without
referring explicitly to topology.

For our analysis
we neglect the reduced speed of light in the optical fiber. Everything the
fiber does, is to guide the light around the wheels. Along the straight lines
the fiber is not necessary. We can thus produce an equivalent situation
by replacing the fiber by stationary mirrors which (disregarding the small
transversal displacement introduced by the wheels) act on the light signals in
the same way. See Fig.~\ref{fig:conveyor}\,(b) for an illustration. The
observer is still moving with $v$.
This situation corresponds to that of Fig.~\ref{fig:conveyor}\,(a) with
infinitely small wheels. In this modified experiment, neither the observer nor
the mirrors are accelerated but we nevertheless detect the Sagnac shift. This
new aspect makes the discussion of this experiment worthwhile.
In contrast to the standard scenario, everything can
also be explained without any difficulties (like time-lags) in the standard
inertial frame of the observer, using standard \emph{continuous global}
Minkowski coordinates. 
In this system the light is being reflected by two uniformly moving mirrors,
which leads to light paths of different lengths, depending on the direction.
This is shown in Fig.~\ref{fig:conveyor}\,(c). The relevant distance between
the mirrors is the distance measured between the events of light
reflection. In the combination of all light paths, this almost trivially leads
to different travel times without being in (not even seemingly) contradiction
to special relativity.
This is no surprise, because the observer moves uniformly along a straight
line and does all the measurement in this global inertial frame.

The more interesting general case, where the fiber moves along any
arbitrarily curved, possibly
self-crossing, path, can be interpreted very easily in terms of
a closed Minkowski space-time whose spatial coordinate is measured along the
fiber. 
It now becomes clear (and was actually tested in the experiment), that the
effect depends only on the length $L$, the speed $v$ and the closed
topology. The specific shape of the fiber in 3-dimensional space is not
relevant at all.

\section{The twin paradox}

The ``twin paradox'' is a second example where principles of relativity
\emph{seem} to be violated if it is discussed without sufficient care.
In fact there is a very close relation to the Sagnac effect as we will see
below. 

\subsection{Time dilation}
\label{sec:time dilation}

Before coming to the twin-paradox itself, we briefly review the standard
meaning of 
time dilation in \sr. Using isotropic synchronization in all inertial frames,
we can apply Lorentz transformations following Eqs.~\eqref{eq:lotra disk
  1}--\eqref{eq:lotra disk 2}. A clock ticking at a fixed position in the
moving system \scom\ has $\diff x'=0$. With this we immediately obtain the
time 
interval $\diff t$ measured in the laboratory frame \slab\ for each interval
of the moving clock $\diff t'$ as
\begin{align}
\diff t &= \gamma \, \diff t' \rtext{.}
\label{eq:time dil 1}
\end{align}
This is the standard form of time dilation. It is very important to
keep in mind the different measurement procedures in both systems and the
very different meaning of the time intervals. In the moving system \scom,
we measure the proper time of a clock along its world line, which is
independent of any coordinate conventions. In the laboratory frame \slab,
on the other hand, we are comparing the readings of clocks at different
positions. We know from the previous discussion that such a measurement
depends on the synchronization convention.
Only this difference allows for the symmetry of time dilation from one system
into another and back. Reading a clock fixed in the laboratory frame \slab\
from the moving frame \scom\ would lead to
\begin{align}
\diff t' &= \gamma \, \diff t \rtext{,}
\label{eq:time dil 2}
\end{align}
without contradiction to Eq.~\eqref{eq:time dil 1}, because the definitions of
$\diff t$ and $\diff t'$ are simply different in both cases.

We could instead have used the alternative convention corresponding to the
global synchronization on the rotating disk from Eqs.~\eqref{eq:lorentz global sync
  first}--\eqref{eq:lorentz global sync last} to find that time is now
\emph{dilated} 
in one direction but \emph{compressed} in the other one,
\begin{align}
 \diff t = \gamma \,\diff T \rtext{,}\equad \diff T = \gamma^{-1}\, \diff t
 \rtext{.} 
\end{align}
An unambiguous comparison of clocks can only be performed when their
corresponding world lines cross.
Such a situation was discussed above in the context of the Sagnac effect,
providing invaluable insights in \sr.
A better known example is the situation leading to the classical twin paradox.

\subsection{Standard situation}

In the standard situation we have two twins, one of whom stays on earth at rest
in the reference frame \slab\ 
$(t,x)$, and the other travels in $x$ direction with a speed $v$. After some
time 
($\Delta t=\tau$ as measured on earth in standard convention), the traveling
twin reverses the direction and returns with speed $-v$.
During this experiment, the sister on earth ages by $2\tau$ and the
traveling twin only by $2\tau'=2\tau/\gamma$, as can be calculated easily. The
difference between $\tau$ and $\tau'$ may seem strange when compared  
with everyday intuition but, since $v$ has to be much larger than normal travel
velocities in order to produce any detectable effects, is not in
contradiction with experience and especially not paradoxical so far.

The paradox becomes apparent only if we try to interpret the same situation
from the reference frame \scom\ of the traveling twin. If the observer on
earth sees 
the traveler's time dilated by a factor $\gamma$, should the same not also be
true in the opposite direction? In other words: Seen from the traveling twin,
her sister on earth seems to travel with speed $-v$ and then $v$, so that (in
this naive and incorrect interpretation) the effect predicted in this
reference frame should be the opposite, $\tau'=\gamma\tau$ instead of
$\tau'=\tau/\gamma$.

The explanation, of course, is that the traveling twin, in order to be able to
return home and compare ages (or clocks), has to accelerate at some point. This
means that there is no \emph{single} inertial system \scom\ associated with
the traveling twin.
When using the general notion of Minkowski space-time and the interpretation of
the $t'$ coordinate as ``time'', the situation is nevertheless strange.
The difference in ages is $\int\!\diff t\,(1-1/\gamma)$ and is thus an integral
over a function which depends on the velocity but not on the
acceleration.\footnote{If we would rewrite the integral to contain the
  acceleration, the integrand would depend on 
the position ($x$ coordinate) or time ($t$ coordinate). Since the
space-time is homogeneous this is highly unphysical.} This means that there
probably is no real physical acceleration effect.

\begin{figure*}[ht]
\includegraphics[width=0.85\textwidth]{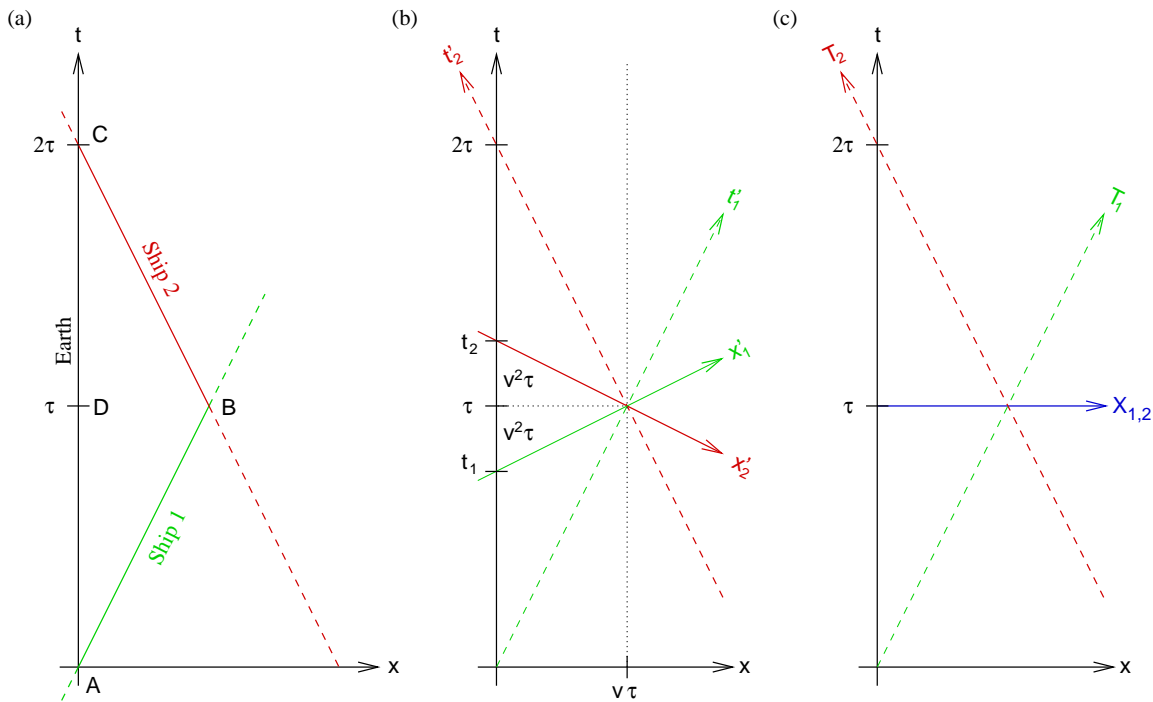}
\caption{(a) Scenario of the modified twin paradox experiment. Clock~0 stays on
  earth ($x=0$) in \slab. Clock~1 travels with spaceship 1 and is synchronized
  with clock~0 
  when passing at $A$. When the two spaceships meet at $B$, clock~2 is
  synchronized to clock~1 and returns to earth. When arriving there at $C$, it
  is  compared with  clock~0.
  (b) Coordinate axes of the moving systems $\scomm_1$ (thin) and $\scomm_2$
  (thick lines). The 
  line $t'_1=0$ intersects $x=0$ at $t=t_1$, the line $t'_2=0$ at $t=t_2$.
(c) Alternative comoving coordinates. Time $T$ is synchronized globally,
  $T=t/\gamma$. The spatial axes $X_1$ and $X_2$ (lines of $T_1=0$ and
  $T_2=0$) coincide when the spaceships meet, so that there is no jump in
  corresponding $t$.}
\label{fig:twins}
\end{figure*}

We can indeed in an equivalent experiment avoid the acceleration
completely, see Fig.~\ref{fig:twins}\,(a). We now have two inertial spaceships,
one traveling with $x=vt$ and the other with $x=v(2\tau - t)$.
When the first spaceship passes the earth at $A$, clock~0 on earth
and clock~1 on the spaceship are both set to zero. The spaceship travels on
and when it meets the second spaceship at $B$, clock~2
on board spaceship 2 is synchronized to the clock~1 on the first
spaceship. Since both clocks pass arbitrarily close to each other, this
synchronization is unique. Now the second spaceship travels on, until it
reaches the earth at $C$ where the clock on earth and on
the spaceship are compared. In this experiment the result will be exactly the
same 
as before but no acceleration effects can be present, because only inertial
world lines and reference frames are considered.
An equivalent thought experiment was proposed by \citet{dingle56} who
unfortunately used incorrect arguments in order
to prove that there is \emph{no} age difference of the twins.

If we call the proper time interval measured by the traveling twin (or on the
two spaceships) $\Delta t'$, we always have $\Delta t=\gamma\Delta t'$. This
should \emph{not} be interpreted naively
in the way that the moving clock is going \emph{more slowly} than the clock on
earth. At each instant we could instead consider the same situation from the
moving system in which we then would say that the moving clock goes
\emph{faster} than the clock on earth. These assertions cannot both be
true. The point here is that $t$ is merely a coordinate which can be defined
according to different synchronization conventions, as shown above.

In the conventional Minkowski coordinates, see Fig.~\ref{fig:twins}\,(b), the
observer on earth sees the 
clocks 1 and 2 being slow by the time dilation factor $\gamma$. Observers on
the spaceships 1 and 2 do see the clock 0 being slow by the same factor.
In the end, both clocks meet again, so that they actually can compare their
readings. The
two notions would therefore be inconsistent if there were no discontinuity at
some point.
We now look at the systems of the spaceships when they meet each
other at $B$, and determine the event $(t,x=0)$ on earth
which is simultaneous (according to isotropic
convention) to this event where the transition from $\scomm_1$ to
$\scomm_2$ takes place. In $\scomm_1$ the simultaneous time on earth is
$t=t_1$, while in $\scomm_2$ it is $t=t_2$. 
This means that when we
change from system 1 to 2 we miss some time interval $t_2-t_1$ of the clock on
earth. This is of course not a real physical effect. When \emph{watching} the
clock on earth from either spaceship, we will not see it instantaneous
but retarded by the light travel time. Only when taking this into
account and correcting for it (using standard convention), do we observe this
jump.
Elementary geometry (see Fig.~\ref{fig:twins}) leads to $t_2-t_1=2\tau
v^2$. The time interval actually taken into account by the traveling twin
as being simultaneous at some instant is $2\tau-(t_2-t_1)=2\tau
(1-v^2)$. Using standard time dilation, the traveling twin sees 
this interval 
dilated by $\gamma$, so that $\tau'=\tau/\gamma$. Taking the time gap
into account, the description in the inertial frame of the traveling twin is
therefore in absolute agreement with the description in the earth's rest
frame, where the time dilation is seen directly and without any coordinate
gaps. 

As in the discussion of the Sagnac effect, there are no problems with
relativistic conventions locally but only globally. Although the moving
systems are inertial on both ways, we cannot combine them to one global
inertial system. With the standard coordinate conventions this inevitably
leads to a discontinuity (the time gap) which introduces the preferred frame
effects. This time gap is equivalent to (twice) the time-lag on the rotating
disk, as we will discuss in more detail below.

\subsection{Discussion in alternative coordinates}

As before we can use a different clock synchronization in order to avoid the
discontinuity. Using the global coordinates introduced in Sec.~\ref{sec:global
  sync}, there is no time jump when reversing the direction of travel, see
Fig.~\ref{fig:twins}\,(c). Now the reference frame of the earth \slab\
\emph{seems} to 
be preferred because we have $\diff T=\gamma^{-1}\,\diff t$ but $\diff
t=\gamma\,\diff T$, so that the transformation is not symmetric anymore. It
can now be said without contradiction that the observer on earth sees the
traveling clocks dilated by 
$\gamma$ but the traveling observer sees clocks on earth going fast by the
same factor.
It is all a matter of convention how the clocks are compared. Sometimes a
non-conventional clock synchronization actually does simplify the discussion
of problems.

\subsection{Twin paradox on rotating disk or in closed space-time}
\label{sec:twins closed}

We found that, in the twin paradox as in the Sagnac effect, global
inconsistencies of standard Minkowski coordinates 
when considering closed world lines can be interpreted without any
acceleration effects and are thus more a topological than a kinematical
problem. To illustrate this even better, let us now consider the twin paradox
in a closed Minkowski space-time which, as we have learned before, is
equivalent to the rim of a (rotating) disk.
This scenario allows the repeated comparison of both twin's clocks (or ages)
even if both are in inertial motion. 

Let us again consider the scenario of Fig.~\ref{fig:twins}.
We can now avoid the acceleration and switching of
reference frames if we change the space-time topology by identifying $x=0$
with $x=L=v\tau$ with equal $t$ coordinates corresponding to each other.
The earth is then at rest in the preferred frame \slab\ of this space-time. 
Measured in \slab, the traveling twin returns to $x=0$ at $B\equiv D$
after the time interval $\tau$ without any acceleration, simply by
surrounding the closed space-time.
Measured by the traveling twin in \scom, the interval is shorter,
$\tau'=\tau/\gamma$, 
although both twins are traveling uniformly. How can this effect now be
explained in the comoving frame \scom?

From the discussion in Sec.~\ref{sec:topology} above we know that the closed
Minkowski space-time is 
equivalent to the situation on the rim of a disk so that both situations can
be analyzed with the same formalism.
The twin at home is at rest on the rim while the traveling twin moves around
once. If we interpret this from the reference frame \scom\ of the traveler,
we have 
a situation equivalent to the Sagnac effect. In this frame we would expect
$\Delta t'=\gamma\tau$, which is not in agreement with the result from the
laboratory frame, $\tau'=\tau/\gamma$. The difference between the two is
the time-lag that was discussed before:
\begin{align}
\Delta t' - \tau' = (\gamma-\gamma^{-1})\,\tau 
= \Theta
\end{align}
Here we used $L'=\gamma L=\gamma v\tau$ to recover the time-lag $\Theta=v L'$
according to Eq.~\eqref{eq:time lag}. The negative sign of Eq.~\eqref{eq:time
  lag appl} has to be taken, because the twin at rest moves in the
\emph{negative} $x'$ direction in \scom.
The time-lag is independent
of the velocity with which the disk is surrounded. One and the same lag is
observed with slow clocks, in the twin paradox and in the Sagnac effect where
light is used. This is in accordance with our interpretation as a topological
effect.

In this scenario, where both twins stay in their inertial frame
for the whole experiment, it is
obvious that acceleration cannot be blamed for the detected age difference or
the asymmetry of time dilation.
This means that, globally, not all inertial frames are equivalent, but there is
a preferred frame which in the case of the disk is given by the non-rotating
laboratory frame and in the case of the closed Minkowski space-time by the
compactification without a time-lag.
We learned that the preferred frame is also the one with minimal length
of the circumference.
The two twins do
indeed age differently, although they are both in global inertial systems which
are equivalent locally. However, this situation is not paradoxical at all,
because in the closed space-time there \emph{is} a unique preferred frame which
accounts for the difference.

The twin paradox in flat closed space-times has been discussed before in
publications by \citetp{brans73}, \citet{low90} and \citetp{dray90}, who
already found that, in addition to the 
local time-dilation effects, global properties of space-time have to be taken
into account, and that indeed these global effects are responsible for the
selection of a preferred frame. Similar effects in closed Minkowski
space-times were discussed by \citetp{peters83}, without explicitly referring
to the Sagnac effect or twin paradox.
The situation was generalized to closed 3+1 space-times by
\citet{barrow01} and \citetp{uzan02}, without changing the conclusions in
so far as they are relevant in our context.
To the knowledge of the author, the equivalence of the twin paradox in open
and closed space-times has 
not been discussed before. This is also true for the exact
equivalence of the 
(rotating) disk with the closed 1+1 Minkowski space-time and some of the
implications for the connection between the Sagnac effect and twin paradox.

\section{Definition and measurement of lengths}
\label{sec:ehrenfest}

In this section we want to discuss the measurement and interpretation of
lengths in connection with different synchronization schemes.
This is of particular interest, because several authors argued that, even in
the rest frame,
length measurements are a matter of convention and do
depend on the adopted synchronization.
We will show that only the measurement of the length of \emph{moving} (with
respect to the observer) objects reflect the conventionality of the
synchronization. 
However, this ambiguity can easily be resolved if the measurement process is
analyzed with some care. We will show that the synchronization which is
relevant for the definition of length is a direct result of how the
measurement is performed.

\subsection{Ehrenfest paradox}

A motivation to study these problems is the discussion of a seeming
paradox which was first presented by \citetp{ehrenfest09}.
For a historical review with discussion of different approaches of solutions,
we refer to the work of \citet{rizzi02} and \citet{groen04} and the
references therein. In the following presentation of the problem, we use
Ehrenfest's original notation, which differs from the rest of this paper.

Consider a rigid cylinder (``ein relativ-starrer Zylinder'')  of radius $R$
and arbitrary height $H$. When at rest, 
the circumference of the cylinder will have a length or $2\pi R$.
The cylinder is now set into stationary rotation around its axis.
The radius of the rotating cylinder measured by an observer at rest is
called $R'$.
We now have two requirements which contradict each other and thus form the
paradox: (a) Seen from the laboratory frame, the circumference moves with some
velocity so that it will be Lorentz contracted. Measured from outside, it will
appear shorter, therefore $2\pi R'<2\pi R$.\footnote{Note that
  with $2\pi R'$ the circumference in the \emph{laboratory} system is
  calculated so that the use of Euclidean geometry is justified. The same is
  true for $2\pi R$ which is the proper circumference of the cylinder,
  measured at rest.}
(b) The radius moves perpendicular to its extension so that its length will
not change, therefore $R'=R$.

The solution of the paradox, as we interpret it, lies in the assumption of the
existence of rigid bodies which can be set in 
rotation. Closely related to this is the validity of Euclidean geometry in the
rotating frame. If space in the
rotating frame is not Euclidean, a rigid body which ``fits'' into Euclidean
space will usually ``not fit'' into Non-Euclidean space. If this is the case,
a real solid body will deform, so that the true effects measured in an
experiment will in addition to relativistic effects also depend on the
elastic properties of the body itself.
In order to avoid this difficulty and to define the problem unambiguously,
we formulate it differently: Consider not a rigid body but the
geometrical space of a rotating disk, defined in the laboratory frame by
$\zeta=0$. For the moment, we do not care how this disk is built. A
solid disk set to rotation may deform, but we assume that we can compensate
this deformation so that we again have a flat platform with $\zeta=0$ on which
we can perform our experiments in the rotating frame \scom.
We now construct on this disk a circle of radius $R'$ (measured in \scom; note
that this definition of $R'$ differs from the one of \citet{ehrenfest09} given
above) 
using a  
collection of standard rods of known length, which are somehow supported in
order to avoid tensions and deformations, and to be sure that they define proper
lengths. This radius will be seen under the same length $R=R'$ in the
laboratory frame because of Ehrenfest's argument (b) above, which is consistent
with relativity.
In the same way we measure the circumference of the circle by laying one rod
behind the other until we return to the first one.
The length measured in this way we call $L'$.
The question is now: How does this comoving length $L'$ relate to the length
$L$ measured in the laboratory frame?
Note the fundamental difference of this scenario from the original
Gedankenexperiment of Ehrenfest. We do not rely on Euclidean geometry and we
do not need rigid bodies. Nevertheless will our discussion explain the
solution of the original paradox.

\nocite{einstein17,einstein31,einstein21,einstein22,einstein56}\citet{einstein17,einstein21,einstein56}
(see also \ocite{einstein11} for early comments on Ehrenfest's publication)
analyses the problem in the following simple way. Seen from the laboratory
frame, we can at a certain time (simultaneous in the standard isotropic sense)
make markings on a reference platform at rest, which is close to the
rotating disk, at the positions where the endpoints of all the measurement
rods on the disk are located at that moment. Then we can count the markings
and measure their distances in the laboratory frame. The number, of course, will
be the same as seen on the disk, but the lengths will appear Lorentz
contracted with the 
usual factor $1/\gamma$. Therefore we see the complete rotating circumference
contracted by the same factor. Measured in the laboratory, the markings will
span the circle with circumference $L=2\pi R$, so that we find $L=L'/\gamma$
or 
\begin{align}
L' &= \gamma L \rtext{,}
\end{align}
exactly as in Eq.~\eqref{eq:L'}.
There is nothing wrong with this reasoning, but since other arguments have
been proposed to ``prove'' different effects, a deeper discussion is definitely
worthwhile and will help in understanding relativity better.
In particular, we will have to discuss the same problem from the perspective of
the rotating reference frame. We found before that the rim of the disk can be
described as a flat 1+1 Minkowski space-time (modulo a global
synchronization oddity). The issue of length on the rim should therefore also
be treatable in the moving frame in terms of \sr.
There is no reason to avoid this discussion with the argument that the system
is accelerated and \sr\ would not be valid here. We can not agree with
\citet{peres04} in this point.

\subsection{Lengths of objects at rest}
\label{sec:lengths rest}

\begin{figure*}[t]
\includegraphics[width=0.75\textwidth]{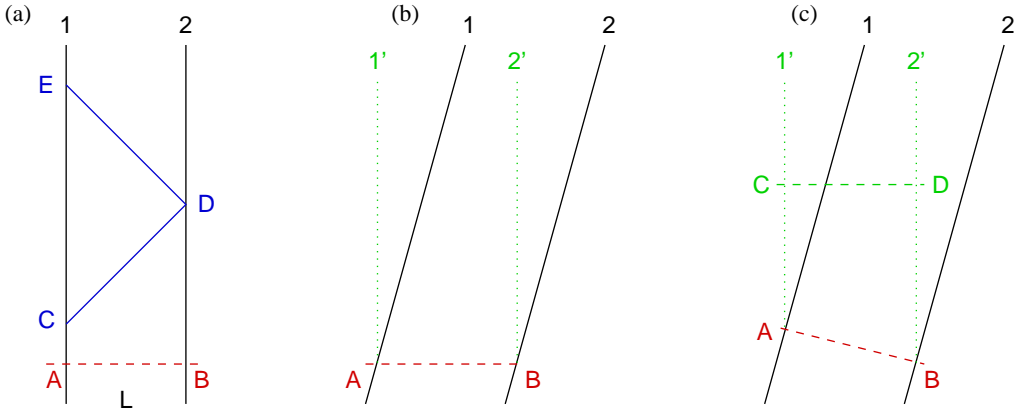}
\caption{(a) Definition of length (or distance) $L$ between two world lines $1$
  and $2$ of particles measured in their common rest frame. We use a light
  signal $CDE$ to measure the distance. The invariant interval along the world
  line of particle $1$ between $C$ and $E$ (proper time measured by $1$) is
  $\Delta s=2L$. An 
  equivalent measure of $L$ is given by the invariant interval along the
  dashed line $AB$, which is defined to be simultaneous to $A$ according to
  standard isotropic synchronization. This interval between $A$ and $B$  is
  $\Delta s^2=-L^2$. 
  These measures of length are independent of coordinates.
(b) Measurement of the distance between moving particles $1$ and $2$. At the
  two standard-simultaneous events $A$ and $B$, markings ($1'$ and $2'$) are
  made. Their distance (equal to the interval $AB$) defines the momentary
  distance of $1$ and $2$.
(c) Distance between moving particles $1$ and $2$ with alternative
  convention. Markings are made at $A$ and $B$ not
  standard-simultaneously. The distance $CD$ (not the interval $AB$) between
  these markings is used as 
  measurement of the distance of $1$ and $2$.
}
\label{fig:length}
\end{figure*}

We have \emph{defined} our general coordinates of Sec.~\ref{sec:coord gen} so
that $X$ measures \emph{proper length} in the rest frame, which is in turn
defined by constant $X$. 
We can therefore start our discussion from this known fact.
It is important to understand that lengths are not measured as intervals
between \emph{events} in space-time, for which an invariant $\diff s^2$
exists, but between \emph{world lines} of objects,
i.e.\ in this case between lines of constant $X_1$ and $X_2$. These lines are
defined 
independent of coordinates but only by the chosen inertial frame.
In order to measure the distance without referring to coordinates, we can
utilize 
the constancy of the two-way speed of light. We send a light signal from
$X_1$ to $X_2$ \emph{and back} and use the invariant proper time $\tau$
measured by a clock at rest at $X_1$ to determine the length or distance
between the two world lines, $L=\tau/2$. The same result would be obtained by
making the measurement at $X_2$ and sending the light to $X_1$ instead.
In our coordinates this leads to $L=|X_1-X_2|$ by construction.
With this measurement procedure at hand, we can now try to interpret the
length geometrically in our 1+1 Minkowski space-time. It is easy to see that
the length equals the integrated invariant intervals in the following way,
\begin{align}
L = \int_{X_1}^{X_2} \diff L = \int_{X_1}^{X_2} \sqrt{-\diff s^2} \rtext{,}
\end{align}
where the integration is performed along a straight line of standard Einstein
simultaneity. See Fig.~\ref{fig:length}\,(a) for an illustration.
This view is confirmed with the general metric of Eq.~\eqref{eq:metric
  general 1}. In 
the isotropic synchronization, we have $A=0$ so that $\diff T=0$ defines a
line of simultaneity. This directly leads to $\diff s^2=-\diff X^2$, i.e.\ the
spatial coordinate really measures length.
However, we should not forget that this convention for $A$ is only one out of
many possibilities. In the general case, a line of standard simultaneity is
defined by $\diff T=A\,\diff X$ so that $\diff s^2$ keeps its invariant value,
just as required.

This shows that, although the clock synchronization is a matter of convention,
the spatial length between two particles at rest is always given by the
invariant interval $\diff s^2$  between the world lines measured along a line
of \emph{standard simultaneity.} Note that (for $A\ne0$) this will generally
\emph{not} be a line of constant $T$.
This geometrical definition of lengths can be used for rest lengths in
\emph{any} inertial  frame without referring to coordinates. In order to
measure the
distance in an experiment, no explicit synchronization has to 
be performed, because the two-way light travel time measured at one position
can be used. This is consistent with the standard convention of \sr\ where
length is measured by the spatial Minkowski coordinates, i.e.\
\emph{orthogonal} to the local time direction.

The definition of lengths is particularly simple in this situation and
should be no matter of debate. Discussions of space (in contrast
to space-time) in more general situations, including accelerations and
even gravitation, were presented by \citet{rizzi02,rizzi04} and
\citet{ruggiero03} and 
in the references therein, in the notion of the \emph{relative
space}. They in a very convincing way use world lines as equivalence 
classes and define the ``relative space'' as quotient space 
relative to these classes. In simple words, a point in relative space is a
world line in space-time which is \emph{defined} to be at rest, or actually
defines the 
local rest frame. This is close to the real experimental situation
where material bodies are used as a reference frame.

According to \ocites{rizzi02,ruggiero03,rizzi04}
(and references by Cattaneo therein), the temporal and spatial projections of
a vector $r^\alpha=(r^T,r^X)$, $\bar r^\alpha$ and $\tilde r^\alpha$,
respectively, 
 are\footnote{See Eq.~\eqref{eq:proj} for the
  definitions.}
\begin{align}
\bar r^\alpha = \begin{pmatrix} r^T-A\,r^X \\ 0 \end{pmatrix} \rtext{,}
\equad
\tilde r^\alpha = \begin{pmatrix} A \\ 1 \end{pmatrix} r^X \rtext{,}
\end{align}
so that the spatial length becomes
\begin{align}
\tilde r^\mu \tilde r_\mu = g_{\mu\nu} \tilde r^\mu \tilde r^\nu 
= -(r^X)^2 \rtext{.}
\end{align}
This confirms that, irrespective of synchronization, the $X$ coordinate
measures length in space, 
in agreement with the definition of the coordinates in Sec.~\ref{sec:length
  X}. This confirms our view which was derived in a more elementary way
above. 
The general definition can be used not only on the rim but on the
complete disk. We do not need this full treatment here and only
mention that, as expected, nothing happens in the radial direction so
that the radius of the disk is the same in both frames.

\subsection{Lengths of moving objects}
\label{sec: moving lengths}

The measurement of lengths of moving objects (or distances of moving
particles) is not as unique as that of 
objects at rest, because we have to define \emph{when} to measure the
positions. This is not a matter of interpretation but of real concrete
physics.
The standard convention of relativity is to use those
events on the two moving particles' world lines which are simultaneous in the
Einstein convention of the observer's frame and take the invariant interval
between them.
A practical measurement process could mark the positions of both moving
particles (or endpoints of a rod) simultaneously in the observer's frame and
then measure the distance 
between these markings with any convenient method.
This is illustrated in Fig.~\ref{fig:length}\,(b) and leads to the
momentary distance, where ``momentary'' has the same ambiguity as
``simultaneous''.
In this standard convention, the measurement $L$ of the length of a moving rod
(proper length $L'$) leads to the well-known Lorentz contraction,
$L=L'/\gamma$.

The use of a different convention of simultaneity, as illustrated in
Fig.~\ref{fig:length}\,(c), will lead to a different distance between two
moving 
particles. The markings are now made at times which are not simultaneous in
the standard convention, so that the distance of the markings differs from the
standard measurement.
The result of a measurement of the length of a moving body does depend on the
exact measurement procedure. The length of a moving object is in the same way
convention dependent as the time difference of clocks at different positions.
This is the core of the ``pole in the barn'' paradox in which a pole,
which at rest does not fit into a barn, is moved with such a high
velocity that its Lorentz contracted version would fit in the
barn. The seeming paradox is that in the system of the barn the pole
now does fit, while in the system of the pole the barn is contracted so
that the pole fits even less than when at rest. ``Fitting in the
barn'' means that at some instant (synchronization dependence here!) both
ends of the pole are located 
within the barn. The difference in the views of both reference frames
lies in the synchronization of time along the pole.

\begin{figure*}[ht]
\includegraphics[width=0.7\textwidth]{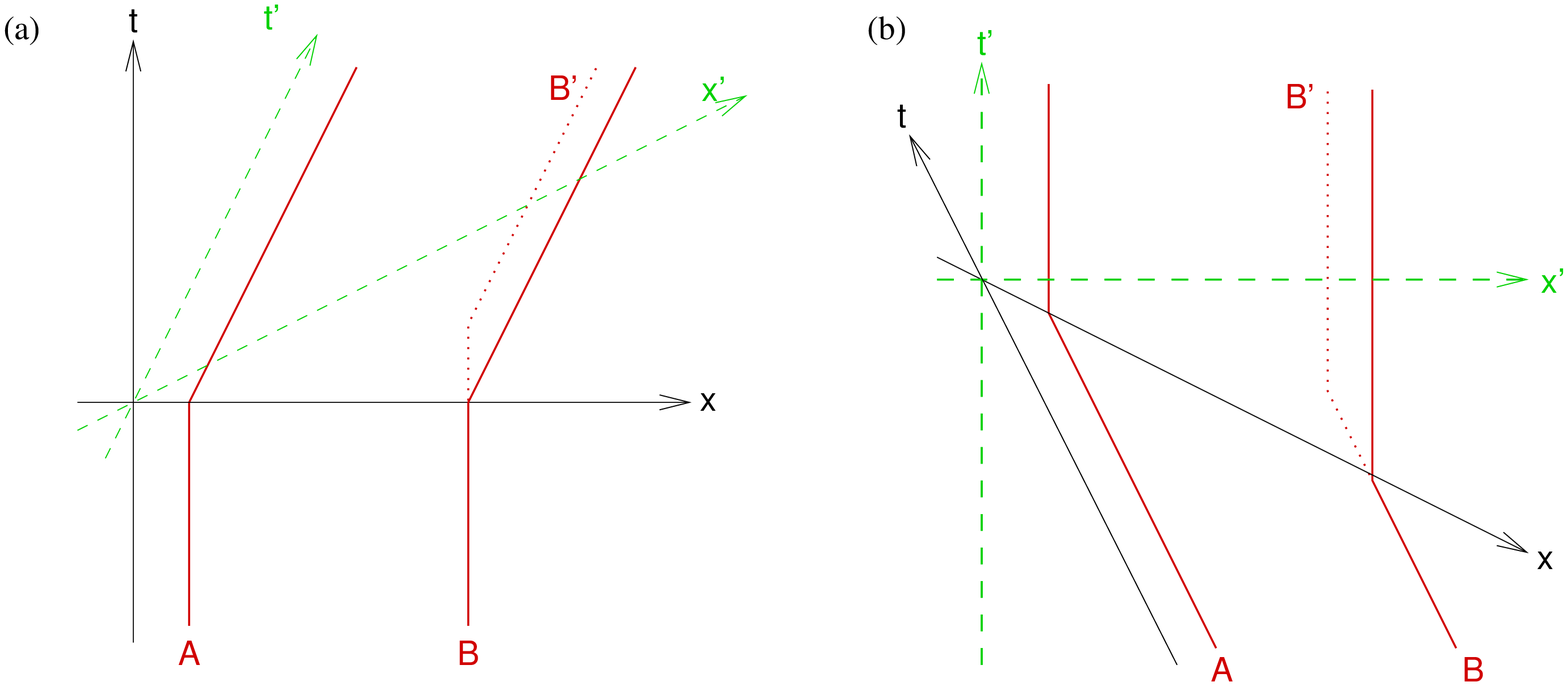}
\caption{Thought experiment discussed by \citet{bell76} (idealized). (a)
  In the   laboratory rest frame \slab, two spacecraft $A$ and $B$ are
  simultaneously 
  accelerated to the same velocity. In their final inertial frame, the
  separation is larger than before the acceleration, so that a thread
  connecting the two before will break. To prevent this, the
  spacecraft have to move closer together, here shown by an
  alternative path for $B'$ which starts its acceleration later. (b)
  Same situation in the final inertial frame \scom\ of the moving
  spacecraft. In this frame the spacecraft do not accelerate
  simultaneously.}
\label{fig:bell}
\end{figure*}

Does this mean that the Lorentz contraction is only a matter of
measuring conventions and is merely illusionary? Not at all, as shown by
a thought experiment which was introduced by \citet{dewan59} and later used by
\citet{bell76,bell87} for a discussion 
of length contractions not only in the Einstein formalism but also in
terms of pre-relativistic Lorentz-FitzGerald contractions.
The idea is illustrated in Fig.~\ref{fig:bell}. Two spacecraft $A$ and
$B$ are at rest in some inertial reference  frame \slab\ at $y=z=0$ with a
distance of 
$L$ along $x$. They are connected with a thin thread of
proper length $L$. Both spacecraft now do accelerate for some time,
using exactly the same program and starting simultaneously in \slab,
so that measured in this rest frame they always have
exactly the same velocity and stay at the same distance $L$.
The question is: Does the thread break? The answer is yes. In
the reference frame \slab, the thread seems to Lorentz-FitzGerald contract
and becomes shorter than $L$. In the comoving frame \scom, on the other hand,
the distance between 
the spaceships does increase, so that the thread of length $L$ cannot span
the distance anymore. An observer in this system would find the reason for the
increasing distance in the fact that the spacecraft do not accelerate
simultaneously. Both observers will see a real dynamical effect, the breaking
of the thread between the spacecraft.
Note that not all authors agree on these conclusions for the experiment. A
recent example is \citetp{field04}, who argues that
the thread will \emph{not} break when accelerating.

\subsection{Discussion of length of rotating circle}

In the frame of the laboratory \slab, the discussion by Einstein presented
above is appropriate. Each element of the circumference can be treated
as a uniformly moving rod which thus contracts. The unambiguously defined
measurement procedure clearly means that standard simultaneity in the
laboratory frame must be used to measure the lengths of the rotating
rim in the laboratory frame. The total length (in contrast to a total
time interval) is simply the integrated local length, so that no
``gaps'' or other effects of the closed topology are to be expected.
This means that the length of the rotating rim $L'$ is observed contracted by a
factor $1/\gamma$ in the laboratory frame \slab. Because we have $L=2\pi R$ in
\slab, this directly leads to $L'=\gamma L=2\pi R\,\gamma$ and
confirms the view which we presented in Sec.~\ref{sec:sagnac minkowski},
Eq.~\eqref{eq:L'}, without discussion. 

We now change to the rotating frame \scom. First we want to show that the
definition of $L'$ really is a measurement of the circumference length
in the rotating frame. The principle of relativity teaches us that the
two-way speed of light is constant, so that we can use two-way light
travel times to measure small elements of length. These can be added
up without problems because, by using the two-way measurement, we are
not affected by the time-lag at all. Even if we really use total round
travel times, the two lags in both directions cancel exactly.
This confirms that the invariant interval measured along a line of standard
simultaneity really leads to the proper length of the rim of the rotating
disk.
It should be noted, however, that the rim of the disk is a closed curve in
\emph{space} but the line corresponding to its length is not closed in
\emph{space-time}. 

In contrast to this, \citet{klauber98} claimed that the length of the rim must
be measured along a 
closed curve in space-time. We have shown that such a measurement would
not be in agreement with relativity.
\citet{klauber98} measures lengths on the rotating disk 
as invariant space-time intervals not along lines of
isotropic simultaneity (``non-time-orthogonal''), but along lines of constant
\emph{global} 
time, i.e.\ in our notation along $\diff T=0$, where the
synchronization is defined as in Sec.~\ref{sec:global sync} with
$A=v$. This leads directly to the absence of Lorentz contraction in
the case of the rotating disk.
It is clear that neither the concept nor the consequence is compatible
with \sr, and indeed the 
theory is meant as a testable alternative to relativity, which moves
the subject outside the scope of this paper.
However, the
idea that the proper length of a moving ruler
somehow depends on global properties of the space-time seems quite
unnatural. It is not clear how this can be compatible with Lorentz contraction
in inertial frames, given the fact that the rim of the disk can locally be
described by inertial frames with unlimited accuracy.
As long as there is no evidence for the failure of \sr\ in
rotating systems, we do not want to discuss alternative theories here.

\begin{figure}[ht]
\includegraphics[width=0.4\textwidth]{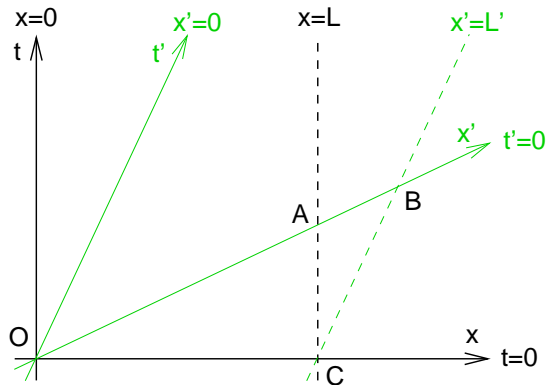}
\caption{Measurement of the length of the non-rotating circle in the
  rotating frame \scom, shown in \slab\ to avoid the time-lag. For the
 laboratory frame \slab, lines of $t=0$ and the 
 start and end of the circle $x=0$ and $x=L$ are shown. The line
 $t'=0$ is according to the isotropic convention in the moving
 frame. ``Endpoints'' of the rotating circle are at $x'=0$ and $x'=L'$.
 The standard measurement along isotropic
  simultaneity uses the invariant interval along $OA$ (leading to
$L/\gamma$) which leaves out
  the part $AB$ of the complete rotating rim $OB$ (corresponding to
 $L'=\gamma L$), which means that the 
 circumference appears contracted without contradiction.
 With global
 synchronization, the interval $OC$ would be measured (corresponding
 to the full circle).} 
\label{fig:rimlength}
\end{figure}

Seen from the laboratory frame \slab, matters looked quite clear, 
but now another seeming
paradox appears. Should the non-rotating circle, which has $L=2\pi R$ in the
laboratory frame, not contract when seen from the moving frame \scom? This
would lead to $\Delta L'=\Delta 
L/\gamma$ and thus \emph{possibly} to the same ratio for the total length,
exactly the opposite of $L'=\gamma L$, which was found before.
That these two views are not really in contradiction is quite easy to
understand. Using the standard procedure of measuring moving lengths (at rest
in \slab) 
in the rotating system \scom, i.e.\ doing local measurements without knowing
about the global nature of the rotation, would definitely lead to exactly the
contraction just described. But does the length of the resting circle $L$
measured in the moving frame \scom\ at all correspond to the total length $L'$
of the moving circle as measured in its rest frame? Naively one could expect
this to be true since the rotating and non-rotating circles seem to fill the
same space.
Nevertheless, the answer is a clear ``no'', as illustrated in
Fig.~\ref{fig:rimlength}. Measuring the resting circle in the moving
frame uses only a fraction $1/\gamma^2$ of the corotating circle of
rods, therefore 
the circle at rests seems to be contracted in the moving frame
($L'/\gamma^2=L/\gamma$). Which 
part of the circle we miss is a matter of where we start the
measurement.
Using a globally synchronized measurement, i.e.\ by marking the
resting rods globally simultaneous in the comoving frame, would measure
a length of $\gamma L$ so that the rim now seems to be \emph{expanded} by
$\gamma$. This is no surprise, because the moving circle and the circle at rest
do, when observed at a line of global simultaneity, fill the same interval in
space-time, so that a consistent ratio must result from the comparison.

The experiment discussed by \citet{bell76} (see Sec.~\ref{sec: moving
  lengths}, especially Fig.~\ref{fig:bell}) can be modified, so that a number
of spacecraft is 
first placed at equal distances on a ring of radius $R$, where they can
slide freely. They are then all accelerated from rest in the same way (as
seen in the rest frame), so that after some time they reach a velocity
of $v$. If they are initially tightly connected by thin threads, these threads
will 
either expand or break, which confirms the result of our discussion.
Neither the fact that the rim of the disk is radially accelerated nor the
closed space-topology do play any
role in the discussion of the Lorentz contraction of its length.

\citet{tartaglia99} claims that no Lorentz contraction
occurs. That work, however, seems to be
based on a circular argument. He starts with a disk at rest, filled with rods
along the rim, which is then set to rotation. He assumes that the radius does
not change, which means he considers the \emph{geometrical} space of the
rotating 
circle. Then he claims that, measured on the rotating disk, the circumference
does not change, because in this system the rods do not contract  ``\dots since
the rulers are laid tail to head \dots'' and their number stays the same. 
However, as pointed out by \citetp{rizzi02}, this is not a consistent approach
because it either assumes that the geometrical length of the circumference
does not change (which is exactly what should be proven with this argument,
i.e.\ circular logic), or the rods will be pulled along the rim so that they
expand under tension, which means they cannot be used as unit length rods
anymore. In this sense the analysis seems to mix the geometrical (here in
radius) with the material situation (here on the circumference). This is very
similar to the original Ehrenfest argument but with the \emph{hidden}
assumption of the existence of a rigid disk.
In the full space-time view, \citet{tartaglia99} says that the length of the
rim measured 
Einstein-simultaneously ($l_0$), which is equal to our $L'$, is not directly
accessible to the moving observer. We have shown how this length can be measured
directly (and as integration of local measurements) using light or rigid
rulers.
Later \citet{tartaglia04b} elaborated in more detail on the argument,
also taking into account criticism of \ocite{rizzi02}. He now uses rods which
are not connected, so that gaps could build up between them if
necessary. The 
argument now is that the space between the rods would behave in the  same way
as the rods themselves so that no gaps could result. However, our discussion of
linear acceleration above has shown that, even in linear motion, the ends of
solid rods behave differently from freely moving particles. When put into
motion, 
the distance of free particles measured in their rest frame
increases, contrary to the behavior of rigid rods. Concerning the 
measurement of the total 2-way round trip time, the author disputes that it
can be used as measurement of length. He uses the argument that the travel
time for light sent from the rim to the center and back would, when measured
on the rim, also be compressed with respect to $R$ so that it does not measure
radial distance, either. This is true, but the
measurement is not done in an appropriate way in this experiment. Two-way
light travel times can be used only \emph{locally} as measurements for
lengths. Tangentially, the situation is homogeneous so that the time intervals
can be added up directly. In the radial direction, however, the situation is
different. Let us assume that the disk is rotating with angular velocity
$\omega$. Lower case letters are used for laboratory inertial coordinates in
\slab. We can 
use the metric of Eq.~\eqref{eq:metric outside}. If we now transform
only spatial coordinates to comoving ones (primed), we get $\phi=\omega
t+\phi'$ and $r'=r$. For purely radial world lines (in \scom,
$\phi'=\text{const}$), the 
metric then reads
\begin{align}
\diff s^2 
= (1-\omega^2r^2)\,\diff t^2 - \diff r^2 \rtext{.}
\label{eq:metric tart 1}
\end{align}
For a local observer comoving with the disk, we have $r=\text{const}$ so that
we obtain for the proper time $\tau$ of this observer
$\diff \tau^2 = (1-\omega^2r^2)\,\diff t^2$.
This means that we can write the metric of Eq.~\eqref{eq:metric tart 1}
locally as 
\begin{align}
\diff s^2&=\diff\tau^2-\diff r^2\rtext{,}
\end{align}
which confirms that light travel times can
be used to measure radial distances \emph{locally.} This argument against the
use of two-way light travel times as length measurements is therefore not
valid.

\section{Discussion}

Although our discussion is about the interpretation of relativity and not
about the outcome of physical experiments, we should keep in mind the
relevance of relativity in rotating systems for our everyday life. One aspect
is the application of the Sagnac effect in technical devices like fiber-optical
gyroscopes, used both in civil navigation and regrettably also in modern
military systems, as discussed by \citetp{stedman97}.
The more fundamental aspect is the fact that earth itself constitutes a
rotating reference frame. Here the situation is much more complicated because
of 
the influence of gravitation and the spheroidal shape of the earth. As a
result of these effects, 
clocks on earth do generally run with different rates. Clocks close to the
equator move faster and therefore run with a slower rate. On the other
hand, clocks at higher altitude should run with a faster rate, as a result of
gravitational time dilation. 
However, for clocks kept at
sea level,\footnote{Sea level is defined as a surface of constant potential, 
  including gravitational and centrifugal contributions.} these two effects
should cancel exactly,
as demanded by the equivalence principle. The earth is not exactly spherical
but slightly oblate, due to the centrifugal acceleration, so that clocks
close to the equator are more distant to the center, which results in the
cancellation of both effects.
We now have the same synchronization problems of
clocks on earth as on the rotating disk. A clock slowly transported around the
earth will desynchronize by some small but measurable amount.  At the equator,
this time-lag is of the order $200\,\text{ns}$, the light travel time for
about $60\,\text{m}$.
In any system of highly accurate time-keeping, this has to be taken into
account, e.g.\ in the global positioning system GPS, as discussed by
\citetp{ashby04}. The official time system cannot be and in fact is not
synchronized according to the Einstein convention.

\subsection{Time}

In the Sagnac effect we observe that, seen from a rotating disk, the round
travel time for light around the rim of the disk depends on the direction.
This non-existence of a \emph{global} universal one-way speed of light $c$ for
round trips 
directly implies that there can be no global time synchronization in which the
speed of light is isotropic. Both properties are equivalent, as shown
by \citetp{minguzzi03b}.

In order to understand the problems of the interpretation of the Sagnac effect
and the Ehrenfest paradox in the context of \sr, we derived the most general
metric which is in agreement with the principles of \sr. We demanded that $X$ measures
length in the rest frame (defined by constant $X$) and that $T$ measures proper
time of clocks at rest in this system. Together with the constancy of the
two-way velocity of light, we arrived at the general metric of
Eq.~\eqref{eq:metric 
  general 1}. This metric describes 
Minkowski space-time in coordinates that are less restricted in 
terms of synchronization of clocks. The desynchronization relative to the
standard isotropic convention of Minkowski coordinates is quantified by the
parameter $A$.
We found that this parameter $A$ is a matter of convention, i.e.\ different
values do not contradict \sr\ but lead to exactly the same physical
results. Nevertheless, the
standard convention, using Einstein synchronization, slow transport of clocks
or other local isotropic methods, is preferred in the way that it leads to
coordinates which reflect the isotropy of space-time.
Generalizations of Minkowski coordinates and Lorentz transformations for
different synchronization conventions have been discussed before, e.g.\
by \citeaut{selleri96},\citepp{selleri96,selleri97} \citetp{minguzzi02},
\citetp{rizzi04b}, and 
\citetp{weber04}, not always
with the  same conclusions as presented in this paper.
In contrast to our work, most of these previous publications start from a
Minkowski reference system and derive the generalized transformations relative
to those coordinates. We instead introduced the generalized metric directly.

Closely related with the synchronization convention is the problem of one-way
velocities of light in opposite directions. In order to define such a one-way
velocity, clock readings at different positions must be compared, so that the
result necessarily depends on the synchronization convention.
If \sr\ is valid, these one-way velocities cannot be measured
unambiguously. If Einstein synchronization is used to set the clocks, the
resulting speed of light must be isotropic by construction. A number of
proposed experiments to detect anisotropies are thus actually comparisons of
different synchronization methods. Since the standard methods have been shown
to be equivalent in \sr, these tests are really tests of the theory of
relativity itself and cannot be used to determine ``the correct''
synchronization parameter or ``the'' one-way velocity of light.
Similar discussions were presented by, e.g., \citetp{vetharaniam93},
\citet{anderson98} and \citetp{croca99}.

In the standard situation of inertial frames in open Minkowski space-time,
this conventionality of synchronization is not a problem. One can simply adopt
the isotropic convention and work in the well-known framework of \sr\ in its
standard formulation. Things are more subtle on the rim of a rotating
disk. A rotating disk is, of course, an accelerated system which can generally
be distinguished from an inertial frame. However, we have shown that as long
as the physical processes are restricted to constant radius $R$, i.e.\ to the
\emph{rim} of a rotating disk, there is no possibility to detect the rotation
by 
purely local measurements. In coordinates appropriate for this situation, the
metric is a standard Minkowski metric, so that the local physics is expected
to be that of an inertial frame. ``Locality'' in this context is an extension
of the standard locality of relativity (discussed, e.g., by
\citeaut{mashhoon90},\citepp{mashhoon90,mashhoon90b,mashhoon04}
\citet{dieks04} and \citet{sorge04}),
where it means that in infinitesimally small regions of space-time, local
comoving 
Minkowski frames approximate the real (possibly curved and accelerated)
space-time arbitrarily well and can thus be used for the local calculations.
In our case the Minkowski metric is valid not only in regions very small
compared to the disk but even in extended regions.

At first view this seems to be in contradiction with the Sagnac effect, where
the total round travel time of light signals, as measured by a comoving
observer, depends on the direction in which the signals travel.
When comparing the local effects (which are not different from
inertial frames) with the global ones, we found that the time coordinate
defined by the standard isotropic synchronization convention can not be used
as a \emph{global} coordinate for the complete rim of the rotating disk
because of a time-lag associated with the round travel. It is this time-lag,
whose sign depends on the direction of travel, which has to be taken into
account when doing global measurements. The time-lag does only depend
on the path along which the synchronization is performed but not on the
particle velocity.
Locally, the speed of light in
standard coordinates is isotropic, but these coordinates do not match
globally, so that the global round travel time is not isotropic anymore.

The non-uniqueness or necessity to introduce a discontinuity in order to use
standard coordinates on the rim of the rotating disk can in
our opinion not be seen as a contradiction but merely as a complication. A
mathematically analogous situation is the complex function 
$f=\ln z$ which with its imaginary part measures the phase of $z$. For
$z=r\exp(\iup \phi)$ we have $\ln z=\ln r + \iup \phi$. This function can be
defined as a continuous and differentiable function in 
any region of $\mathbb{C}$ as long as a cut is made along any line going from
zero to infinity. This cut is necessary to account for the ambiguity of the
phase $\phi$. When crossing this line in either direction, the phase jumps by
$\pm2\pi$ so that $\ln z$ makes a jump as well. Without the cut, analytical
continuation around zero would lead to different values, depending on the path.
This is the same behavior as that of Minkowski time on the rim of the
rotating disk. The function is well behaved locally but becomes ambiguous
or discontinuous globally.

Another example is the measurement of time on earth. This analogy does indeed
go further than mentioned by \citetp{bergia98}.
Civil time is based on the mean local solar time, which in a simplified notion
is the time measured by a sundial.\footnote{We
  neglect the discrete time-zones of full or half hours in this discussion and
  assume that the real local solar time is used.} This time coordinate
conforms with our 
demand that $T$ measures the proper time of a local clock. The synchronization
is performed indirectly either by using sundials or in some equivalent
way. This defines
a very convenient time system, although it shows the same non-unique behavior
as the standard Minkowski time on the rotating disk. Let us  imagine that
at some fixed Greenwich mean time GMT (which corresponds to the globally
synchronized time) we shift our position eastwards. For each $15^\circ$ of
longitude, the 
local time will advance by one hour. When going around completely, local time
has advanced by 24 hours. Taking the date into account, this convention means
that a fixed clock will be desynchronized with itself by
one day. In order to avoid this, a discontinuity, defined by the international
date-line, must be introduced into the system. When
crossing this line eastwards, the clocks have to be set back by one day. In
this way the time coordinate stays unique but is not continuous anymore.
Nevertheless, the local solar time is a much better
time coordinate for most everyday life  than the global GMT.
Even the difference between global round travel times and local coordinate
velocities can be seen quite well in this example.
When Phileas Fogg and Passepartout travel \emph{around the
world in 80 days,}\citepp{verne1873} they go eastwards with a mean coordinate
velocity of 
$360^\circ/81\,\text{days}$.  Nevertheless, they return after a time interval
of only 80~days, measured locally by a fixed clock in London.
If instead Phileas Fogg and Passepartout had parted, one traveling
eastwards and one westwards, both with the same local coordinate
velocity, their round trip times measured in London would have been 79 and
81~days, respectively. This 
difference of 2~days corresponds to the Sagnac effect.
The physical reason, of course, is very different in this situation. Local
solar time is \emph{not} what is measured by a slowly transported clock
(unless it is a sundial), and
the coordinate velocity defined in this way is not directly related to
momentum or locally measured speed. In addition, nobody would call twelve at
noon in London ``simultaneous'' to twelve at noon in Yokohama.
Still the similarity is quite striking.

In the specific situation of the rotating disk, an alternative convention of
synchronization can be 
more convenient to describe the physics. We have shown (in agreement with
e.g.\ \citeaut{selleri96},\citepp{selleri96,selleri97}
\citet{goy96c,goy97c,goy97d}
but also with \citet{klauber98,klauber03})
that the special choice of the synchronization parameter
$A=v$ leads to a \emph{globally} valid time coordinate $T$. Using this
coordinate system, the description of local effects is in agreement with
global effects. We have to remember, however, that this alternative
description is not in 
disagreement with \sr\ and does not lead to measurable physical effects
different from the standard convention. It is merely a matter of convenience
(or taste or philosophy) which coordinates are used.

Not all authors do agree on the conventionality of synchronization, in some
cases because the difference between local and global effects is not being
recognized. 
In a number of publications, the circumference length is divided by the round
travel time to 
derive a ``global velocity of light'' which is then clearly anisotropic.
However, in relativity velocities are defined locally as vectors
in the local tangential space. If global measurements
are used to derive velocities, one has to study how the local effects add up
to become global effects. We have shown how this leads to time-lags on the
rotating disk, which are in agreement with the anisotropic global behavior.

\citet{selleri96,selleri97,selleri04} discusses alternative synchronization
conventions in a very specific way. He starts by deriving the most general
transformation, admitting that \emph{a priori} any convention can
be used. His free parameter $e_1$ is related to our $A$ by the simple
equation $e_1=(A-v)\gamma$.
In the case of the rotating disk, however, the author does not cleanly
distinguish between local and global effects. He claims that, due to symmetry,
the globally anisotropic speed of light must necessarily also lead to a local
anisotropy so that global synchronization \emph{must} be used. This would only
be 
true in the absence of any time-lag. Our 
discussion shows, that isotropic coordinates can explain the global anisotropy,
if the time-lag is taken into account.
\citet{selleri96,selleri97,selleri04} claims that the ``global'' or
``absolute'' synchronization ($e_1=0$, $A=v$),
which leads to what he calls ``inertial transformations'', is more natural than
the special relativistic convention. In order to support this view he
additionally proposed an example with linear motion.
The situation is very similar to the experiment we discussed in
Sec.~\ref{sec: moving lengths}, see Fig.~\ref{fig:bell}. On the two spaceships
we now have one clock each which are initially synchronized in the laboratory
frame \slab, using the standard convention. Even after the acceleration stops,
they will 
remain to be synchronized in this system. Seen in the comoving inertial frame
\scom, 
however, the synchronization will violate the standard Einstein convention.
The author uses this as an argument for the existence of absolute simultaneity
and thus for the global synchronization.
We do not find this very convincing, because this kind of synchronization
can be achieved only by first synchronizing in the
laboratory frame or by later 
somehow referring to that frame. It is no surprise that if we use \slab\
in order to synchronize clocks, they will be synchronized in
exactly this frame.
With measurements in the moving inertial
frame \scom, this ``absolute'' synchronization could not be defined. Actually,
the 
system is exactly equivalent to the laboratory frame, so that the same
conventions can be used in both.
We emphasize that we do not argue against the \emph{possibility} of using this
global synchronization convention. The discussion of
\citet{selleri96,selleri97,selleri04} 
indeed shows very convincingly that this can be done and is in certain cases
more convenient than the standard convention.
We merely do not accept the absolute nature of this convention, nor the claim
that \sr\ cannot explain the Sagnac effect.
If any convention plays a special role in a given inertial frame,
without referring to other frames, it is the Einstein convention because
of its isotropic nature. This still holds for the rotating disk as long as no
global properties are probed.

Arguments similar to Selleri's were presented by \citet{goy97c} and
\citetp{goy97d}, who 
argue that if the clocks were synchronized in the beginning, they can for
symmetry reasons not desynchronize when set to (circular) motion. It is true
that clocks stay synchronized according to some convention, but this
convention refers to the system \slab, so that it is in no way absolute for
all systems. Relative to Einstein synchronization in \scom, the clocks
\emph{do} desynchronize, but this is not violating homogeneity. No position is
preferred but only directions. The desynchronization is homogeneous but
not isotropic, and the anisotropy is defined by a real physical process,
namely the acceleration in a certain direction.

Views similar to those of \citet{selleri96} and \citetp{goy97c}, concerning the
``absolute'' nature 
of the global synchronization on the rotating disk, were presented by
\citeaut{klauber98}.\citepp{klauber98,klauber03} Again we feel that a more
thorough distinction 
between local and global effects would lead to a deeper understanding.

Concerning the interpretation of the Sagnac
effect, our work agrees with many other publications, e.g.\ by
\citetp{bergia98}, \citeaut{rizzi98},\citepp{rizzi98,rizzi99}
\citetp{pascual04}, 
\citetp{rizzi04b}, \citet{dieks04} and \citetp{weber04}.
We emphasize, however, that our discussion completely avoids effects of
acceleration. In our formulation, the rim of a rotating disk is treated as an
inertial frame, not only in \emph{infinitely} small regions (tangential space),
but over \emph{extended} regions, and in fact even everywhere if the topology
is taken into account appropriately. This contrasts with and thus extends the
work of \citetp{rizzi04b}, who explicitly use local comoving inertial frames
only for infinitesimally small regions, and others. The approach of
\citetp{selleri97}, who 
considers the limit of $R\to\infty$ with constant $v$ and nevertheless finds a
constant ratio of global round travel times $\tau_-/\tau_+$, interpreted as
ratio of one-way velocities of light $c_+/c_-$, is very
instructive but not really necessary in view of our discussion. 

In order to make the difference between local and global effects even clearer,
we 
introduced a spatially closed Minkowski space-time which was shown to be
equivalent to the space-time on the rim of a disk. In this space-time, there is
\emph{by construction} no acceleration, so that all local effects (measured by
purely local means) are exactly the same as in a standard Minkowski
space-time.
Globally, the rotation velocity is defined by the offset in time introduced
when glueing the space-time together. Local effects cannot be influenced by
this offset, which confirms our view that there can be no objective local
anisotropic effects, neither in this space-time nor on the rotating disk.
Anisotropy becomes detectable only when world lines go around the 
closed space-time (or disk) completely.
It is now a matter of convention to say whether the preferred frame
effects are only a result of the periodic boundary conditions, or if they are
``in reality'' local but can not be detected in an open topology. We prefer
the former interpretation, which avoids the introduction of aspects of reality
which cannot be detected \emph{by principle}.

The fiber optic conveyor experiment shows the
topological nature of the Sagnac effect in a very nice way. We discussed that
here it is explicitly shown that the observed effect depends only on the
closed path itself and on the velocity and the length of the light paths, but
not on the specific shape or on the
acceleration along the path. One might imagine the closed Minkowski space-time
of Fig.~\ref{fig:closed mink} embedded in a higher dimensional space not as a
cylinder but following the shape of the fiber in the experiment. Still it is
only the topology, the speed and total length which define the observable
effect.

We found a close relation between the Sagnac effect and the twin paradox,
especially when the latter is discussed in a closed Minkowski space-time or on
a rotating disk. The paradoxical asymmetry between the two twins is exactly
given by the time-lag observed when surrounding the rotating disk. Since both
twins move along lines which are not topologically equivalent (homotopic),
their proper time experiences different time-lags in comparison to a locally
measured coordinate time.
We found that the twin being at rest in the frame in which the disk is not
rotating ages most quickly. This is not paradoxical, because not all inertial
frames are equivalent in the closed topology case. The frame corresponding to
the non-rotating disk is also the one with the smallest circumference
length. This holds both on the rim of the disk and in the closed Minkowski
space-time. In this way the twin-paradox can be explained as a topological
effect as well.

Similar non-trivial effects imposed by topology are discussed in the context
of cosmology by \citet{barrow03} and \citet{uzan02}\phantom{.} These authors 
show that, if the Universe is spatially closed, the preferred frame induced by
this topology is the same as the preferred frame defined by the cosmic
expansion.

It is somewhat ironic, that in relativistic physics global measurements on the
rim of a rotating disk can detect rotation with respect to an inertial frame,
which is not possible in non-relativistic physics.\footnote{Note that Foucault's
pendulum does not work on the rim of a rotating disk because it can detect the
rotation only if it is allowed to move freely in the radial direction.}
When sending test particles
in both directions with the same velocity (measured in the rotating frame),
they would take equal times for their round trips in Galilean physics, in
contrast to the 
relativistic case, where they experience a time-lag which does not depend on
their velocity but only on the speed of rotation and the length of the rim of
the disk. The situation would be unchanged for a closed space(-time) in both
cases. 
In this sense, motion in relativity seems to be ``less relative'' than in
non-relativistic physics.
In this context it is interesting to note that the situation is very different
in quantum mechanics, where rotation can be detected
even in the nonrelativistic limit (Schr\"odinger equation). This is a hint for
a close connection between special relativity and quantum mechanics.
We refer to
\citetp{anandan81}, \citetp{anandan04}, \citet{dieks91} and \citet{rizzi04}
for discussions of the Sagnac effect in quantum mechanics.

\subsection{Space}

In standard relativity, proper lengths of bodies at rest are
directly given as differences of their spatial Minkowski
coordinates. Formulated in a coordinate independent way, this means that
lengths are measured as invariant intervals along lines of standard
simultaneity, defined in the rest frame.
In contrast to this, lengths of \emph{moving} bodies are not defined
uniquely. The length of a rod is defined as the distance between the world
lines of its ends. The word ``distance'' can be translated as invariant
interval, but then it must be defined along which line this interval is
measured. This means it must be clear, at which time both ends of the rod are
taken. 
This close relation to the conventions of simultaneity was our motivation to
study lengths and especially the Ehrenfest paradox in some detail. The
standard convention of taking isotropic simultaneity in the inertial frame
where the length is to be measured leads to the well known effect of Lorentz
contraction in its standard form. Generally, however, the length of a moving
object depends on the exact measurement process applied in the experiment.

On the rotating disk, the measurements of lengths are less ambiguous than the
measurements of time, so that the Ehrenfest paradox can be discussed without
difficulty.
When measuring the length of the rotating circle ($L'$ proper length measured
in the moving frame \scom) in the laboratory frame, we
would typically compare it with the resting circle's circumference, which has a
proper length of $L=2\pi R$ in the laboratory frame \slab. In order to measure
the 
complete rotating circle at once (without missing parts or counting parts
doubly), standard simultaneity in \slab\ must be used. This is 
equivalent to taking a photograph from above the disk, so that the arguments
by   
\citet{einstein17,einstein21,einstein56} lead to the correct result. The rim
of the disk is observed contracted by $1/\gamma$, so that $L=L'/\gamma$. This
in turn means that the length of the corotating rim $L'$ is \emph{larger} than
$2\pi R$ by a factor of $\gamma$.
This also confirms the original argument of \citetp{ehrenfest09}.

In order to avoid the notion of rigid bodies, we discussed the geometrical
space defining a rotating disk. We therefore find that in the comoving frame
\scom, the circumference of the circle is larger than in Euclidean
space. This means that the rotating geometrical disk
exhibits a 
negative curvature. A rigid disk or cylinder can therefore not be put from
rest (Euclidean geometry) to rotation (curved space).

What happens to a real solid disk when it is set to rotation, depends on
its elastic properties 
and other details of the experiment. Several simple limiting cases shall be
discussed.
(a) The spokes of the disk are assumed to stay rigid but the rubber
circumference is allowed to expand. In this way the result would be equivalent
to the discussion of the geometrical disk. The corotating circumference is
expanded, so that its Lorentz contracted version observed in the laboratory
corresponds to the circle at rest. The observer in \slab\ would not see any
change of the circle, because the elastic expansion and the Lorentz
contraction cancel.
(b) The spokes are made of rubber but the circumference keeps its length. In
this case the radius will shrink in order to decrease the geometrical
circumference length. This contraction could be observed from
\slab\ and \scom.
The cases (a) and (b) are discussed in a recent discussion by
\citet{davidovic04} under the names of ``star disc'' and ``ring disc'',
respectively.

A third special case is also of interest:
(c) The disk is rigid along its two-dimensional extension but flexible in the
$\zeta$ direction. When set to rotation, it would then have to bend in the
vertical direction so that the positive curvature obtained in this way
compensates for the negative curvature caused by the rotation. The surface
would seem positively curved from \slab, but intrinsically
flat  when observed from \scom.

This discussion shows that the Ehrenfest paradox is real and can only be
solved by understanding that rigid bodies are not consistent
with \sr, or at least they cannot accelerate arbitrarily.
Linear accelerations are not a problem, because the Lorentz contraction can
``borrow'' space from outside. On the rotating disk or in the closed
Minkowski space-time, this is not possible because of the restrictions imposed
by the topology.

Our analysis confirms the original arguments of \citet{ehrenfest09} and
\citeaut{einstein17},\citepp{einstein17,einstein21,einstein56} and showed that
an  
explanation consistent with those views can also be derived in the corotating
frame without any difficulty. We therefore do not agree with the views of 
\citeaut{tartaglia99},\citepp{tartaglia99,tartaglia04b}
\citet{klauber98} and \citetp{peres04}, who claim that there is no Lorentz
contraction and thus no Ehrenfest paradox.
On the other hand, our discussion is in good agreement with the work of various
other authors, e.g.\ 
\citetp{rizzi02}, \citetp{ruggiero03}, \citet{dieks04} and \citetp{weber04}.

\section{Conclusions}

All the discussions in this paper have in common the importance of clear
definitions of the concepts used. Once \emph{time} and \emph{length/space}
are defined unambiguously, the measurements in all the (real and thought)
experiments can be translated relatively easily into their corresponding
space-time concepts.
The rim of a rotating disk can be treated as an inertial system without any
contradiction, as long as the radial coordinate is not probed. Unless the
disk is surrounded, the situation is exactly equivalent to that of linear
motion as discussed in standard relativity textbooks. When the global
structure, i.e.\ 
the topology, of the rotating circle becomes relevant, one has to ensure that
the locally valid coordinates and space-time descriptions really match or
``mesh'' 
globally. On the rotating disk, we found that local Minkowski coordinates can
not be extended globally,  which gives the explanation for the Sagnac effect.
In order to overcome any lack of confidence in our notion that the
acceleration is not important for the 
discussion of effects on the rim of the rotating disk, we showed the
equivalence with the situation of a topologically closed Minkowski space-time,
which can physically not be distinguished from the rim of a rotating disk.
We have to keep in mind that the standard notion of special relativity using
Lorentz 
transformations leads to coordinates which are valid locally. In standard open
space-time situations, these coordinates can be extended to form globally
valid systems. Periodic boundary conditions or closed space-time topology, on
the other hand, will restrict the allowed range of these
coordinates. Additional discontinuities or time-lags can appear when matching
the local coordinates globally.
This leads to preferred frame effects which are purely global and of
topological nature.
Although most fundamental physics is usually formulated in terms of partial
differential equations, we have to keep in mind that such a local description
can naturally not explain the world completely. Boundary conditions and thus
the topology of space-time can play an equally important role.

\newcommand{\mnras}{Mon.\ Not.\ R.\ Astron.\ Soc.}
\newcommand{\pz}{Phys.\ Z.}
\newcommand{\fp}{Found.\ Phys.}
\newcommand{\fpl}{Found.\ Phys.\ Lett.}
\newcommand{\physrep}{Phys.\ Rep.}
\newcommand{\grg}{Gen.\ Relativ.\ Gravit.}
\newcommand{\ajp}{Am.\ J.\ Phys.}
\newcommand{\ejp}{Europ.\ J.\ Phys.}
\newcommand{\cqg}{Class.\ Quant.\ Gravit.}
\newcommand{\grqc}{ArXiv Gen.\ Relat.\ Quant.\ Cosmol.\ e-print}
\newcommand{\physics}{ArXiv Phys.\ e-print}
\newcommand{\ncim}{Nuovo Cimento}
\newcommand{\rpp}{Rep.\ Prog.\ Phys.}
\newcommand{\ppsa}{Proc.\ Phys.\ Soc.\ London, Sect.~A}
\newcommand{\comptrend}{C.\ R.\ Acad.\ Sci.}
\newcommand{\jdp}{J.\ Phys.\ (Paris)}
\newcommand{\annphys}{Ann.\ Phys.\ (Leipzig)}

\let\cite=\ocite


\begin{thebibliography}{80}
\expandafter\ifx\csname natexlab\endcsname\relax\def\natexlab#1{#1}\fi
\expandafter\ifx\csname bibnamefont\endcsname\relax
  \def\bibnamefont#1{#1}\fi
\expandafter\ifx\csname bibfnamefont\endcsname\relax
  \def\bibfnamefont#1{#1}\fi
\expandafter\ifx\csname citenamefont\endcsname\relax
  \def\citenamefont#1{#1}\fi
\expandafter\ifx\csname url\endcsname\relax
  \def\url#1{\texttt{#1}}\fi
\expandafter\ifx\csname urlprefix\endcsname\relax\def\urlprefix{URL }\fi
\providecommand{\bibinfo}[2]{#2}
\providecommand{\eprint}[2][]{\url{#2}}

\bibitem[{\citenamefont{Reichenbach}(1924)}]{reichenbach24}
\bibinfo{author}{\bibfnamefont{H.}~\bibnamefont{Reichenbach}},
  \emph{\bibinfo{title}{{Axiomatik der relativistischen Raum-Zeit-Lehre}}}
  (\bibinfo{publisher}{Vieweg, Braunschweig}, \bibinfo{year}{1924}),
  \bibinfo{note}{english version in \ocite{reichenbach69}}.

\bibitem[{\citenamefont{Reichenbach}(1969)}]{reichenbach69}
\bibinfo{author}{\bibfnamefont{H.}~\bibnamefont{Reichenbach}},
  \emph{\bibinfo{title}{{Axiomatization of the theory of relativity}}}
  (\bibinfo{publisher}{Berkeley University Press}, \bibinfo{year}{1969}),
  \bibinfo{note}{english version of \ocite{reichenbach24}}.

\bibitem[{\citenamefont{{Post}}(1967)}]{post67}
\bibinfo{author}{\bibfnamefont{E.~J.} \bibnamefont{{Post}}},
  \bibinfo{journal}{\rmp} \textbf{\bibinfo{volume}{39}}, \bibinfo{pages}{475}
  (\bibinfo{year}{1967}).

\bibitem[{\citenamefont{{Hasselbach} and {Nicklaus}}(1993)}]{hasselbach93}
\bibinfo{author}{\bibfnamefont{F.}~\bibnamefont{{Hasselbach}}}
  \bibnamefont{and}
  \bibinfo{author}{\bibfnamefont{M.}~\bibnamefont{{Nicklaus}}},
  \bibinfo{journal}{\pra} \textbf{\bibinfo{volume}{48}}, \bibinfo{pages}{143}
  (\bibinfo{year}{1993}).

\bibitem[{\citenamefont{{Stedman}}(1997)}]{stedman97}
\bibinfo{author}{\bibfnamefont{G.~E.} \bibnamefont{{Stedman}}},
  \bibinfo{journal}{\rpp} \textbf{\bibinfo{volume}{60}}, \bibinfo{pages}{615}
  (\bibinfo{year}{1997}).

\bibitem[{\citenamefont{Anderson et~al.}(1998)\citenamefont{Anderson,
  Vetharaniam, and Stedman}}]{anderson98}
\bibinfo{author}{\bibfnamefont{R.}~\bibnamefont{Anderson}},
  \bibinfo{author}{\bibfnamefont{I.}~\bibnamefont{Vetharaniam}},
  \bibnamefont{and} \bibinfo{author}{\bibfnamefont{G.~E.}
  \bibnamefont{Stedman}}, \bibinfo{journal}{\physrep}
  \textbf{\bibinfo{volume}{295}}, \bibinfo{pages}{93} (\bibinfo{year}{1998}).

\bibitem[{\citenamefont{Rizzi and Ruggiero}(2004{\natexlab{a}})}]{rrf04}
\bibinfo{editor}{\bibfnamefont{G.}~\bibnamefont{Rizzi}} \bibnamefont{and}
  \bibinfo{editor}{\bibfnamefont{M.}~\bibnamefont{Ruggiero}}, eds.,
  \emph{\bibinfo{title}{Relativity in Rotating Frames}}
  (\bibinfo{publisher}{Kluwer}, \bibinfo{year}{2004}{\natexlab{a}}).

\bibitem[{\citenamefont{Rizzi and Ruggiero}(2004{\natexlab{b}})}]{rizzi04}
\bibinfo{author}{\bibfnamefont{G.}~\bibnamefont{Rizzi}} \bibnamefont{and}
  \bibinfo{author}{\bibfnamefont{M.~L.} \bibnamefont{Ruggiero}},
  \emph{\bibinfo{title}{The relativistic Sagnac effect: two derivations}}, in
  \cite{rrf04} (\bibinfo{year}{2004}{\natexlab{b}}), \eprint{gr-qc/0305084}.

\bibitem[{\citenamefont{Sagnac}(1913{\natexlab{a}})}]{sagnac13a}
\bibinfo{author}{\bibfnamefont{G.}~\bibnamefont{Sagnac}},
  \bibinfo{journal}{\comptrend} \textbf{\bibinfo{volume}{157}},
  \bibinfo{pages}{708} (\bibinfo{year}{1913}{\natexlab{a}}).

\bibitem[{\citenamefont{Sagnac}(1913{\natexlab{b}})}]{sagnac13b}
\bibinfo{author}{\bibfnamefont{G.}~\bibnamefont{Sagnac}},
  \bibinfo{journal}{\comptrend} \textbf{\bibinfo{volume}{157}},
  \bibinfo{pages}{1410} (\bibinfo{year}{1913}{\natexlab{b}}).

\bibitem[{\citenamefont{Sagnac}(1914)}]{sagnac14}
\bibinfo{author}{\bibfnamefont{G.}~\bibnamefont{Sagnac}},
  \bibinfo{journal}{\jdp} \textbf{\bibinfo{volume}{4}}, \bibinfo{pages}{177}
  (\bibinfo{year}{1914}).

\bibitem[{\citenamefont{Pascual-S{\'a}nchez
  et~al.}(2004)\citenamefont{Pascual-S{\'a}nchez, San~Miguel, and
  Vicente}}]{pascual04}
\bibinfo{author}{\bibfnamefont{J.-F.} \bibnamefont{Pascual-S{\'a}nchez}},
  \bibinfo{author}{\bibfnamefont{A.}~\bibnamefont{San~Miguel}},
  \bibnamefont{and} \bibinfo{author}{\bibfnamefont{F.}~\bibnamefont{Vicente}},
  \emph{\bibinfo{title}{Isotropy of the velocity of light and the Sagnac
  effect}}, in  \cite{rrf04} (\bibinfo{year}{2004}).

\bibitem[{\citenamefont{Dieks}(2004)}]{dieks04}
\bibinfo{author}{\bibfnamefont{D.}~\bibnamefont{Dieks}},
  \emph{\bibinfo{title}{Space and Time in a Rotating World}}, in  \cite{rrf04}
  (\bibinfo{year}{2004}).

\bibitem[{\citenamefont{Mashhoon}(1990{\natexlab{a}})}]{mashhoon90}
\bibinfo{author}{\bibfnamefont{B.}~\bibnamefont{Mashhoon}},
  \bibinfo{journal}{\pl} \textbf{\bibinfo{volume}{A~143}}, \bibinfo{pages}{176}
  (\bibinfo{year}{1990}{\natexlab{a}}).

\bibitem[{\citenamefont{Mashhoon}(1990{\natexlab{b}})}]{mashhoon90b}
\bibinfo{author}{\bibfnamefont{B.}~\bibnamefont{Mashhoon}},
  \bibinfo{journal}{\pl} \textbf{\bibinfo{volume}{A~145}}, \bibinfo{pages}{147}
  (\bibinfo{year}{1990}{\natexlab{b}}).

\bibitem[{\citenamefont{Mashhoon}(2004)}]{mashhoon04}
\bibinfo{author}{\bibfnamefont{B.}~\bibnamefont{Mashhoon}},
  \emph{\bibinfo{title}{The hypothesis of locality and its limitations}}, in
  \cite{rrf04} (\bibinfo{year}{2004}).

\bibitem[{\citenamefont{Sorge}(2004)}]{sorge04}
\bibinfo{author}{\bibfnamefont{F.}~\bibnamefont{Sorge}},
  \emph{\bibinfo{title}{Local and global anisotropy in the speed of light}}, in
   \cite{rrf04} (\bibinfo{year}{2004}).

\bibitem[{\citenamefont{Reichenbach}(1928)}]{reichenbach28}
\bibinfo{author}{\bibfnamefont{H.}~\bibnamefont{Reichenbach}},
  \emph{\bibinfo{title}{Philosophie der Raum-Zeit-Lehre}}
  (\bibinfo{publisher}{de Gruyter, Berlin \& Leipzig}, \bibinfo{year}{1928}),
  \bibinfo{note}{english version in \cite{reichenbach58}}.

\bibitem[{\citenamefont{Reichenbach}(1958)}]{reichenbach58}
\bibinfo{author}{\bibfnamefont{H.}~\bibnamefont{Reichenbach}},
  \emph{\bibinfo{title}{The philosophy of space \& time}}
  (\bibinfo{publisher}{Dover, New York}, \bibinfo{year}{1958}),
  \bibinfo{note}{english version of \cite{reichenbach28}}.

\bibitem[{\citenamefont{{Dieks}}(1991)}]{dieks91}
\bibinfo{author}{\bibfnamefont{D.}~\bibnamefont{{Dieks}}},
  \bibinfo{journal}{\ejp} \textbf{\bibinfo{volume}{12}}, \bibinfo{pages}{253}
  (\bibinfo{year}{1991}).

\bibitem[{\citenamefont{{Robertson}}(1949)}]{robertson49}
\bibinfo{author}{\bibfnamefont{H.~P.} \bibnamefont{{Robertson}}},
  \bibinfo{journal}{\rmp} \textbf{\bibinfo{volume}{21}}, \bibinfo{pages}{378}
  (\bibinfo{year}{1949}).

\bibitem[{\citenamefont{{Mansouri} and
  {Sexl}}(1977{\natexlab{a}})}]{mansouri77a}
\bibinfo{author}{\bibfnamefont{R.}~\bibnamefont{{Mansouri}}} \bibnamefont{and}
  \bibinfo{author}{\bibfnamefont{R.~U.} \bibnamefont{{Sexl}}},
  \bibinfo{journal}{\grg} \textbf{\bibinfo{volume}{8}}, \bibinfo{pages}{497}
  (\bibinfo{year}{1977}{\natexlab{a}}).

\bibitem[{\citenamefont{{Mansouri} and
  {Sexl}}(1977{\natexlab{b}})}]{mansouri77b}
\bibinfo{author}{\bibfnamefont{R.}~\bibnamefont{{Mansouri}}} \bibnamefont{and}
  \bibinfo{author}{\bibfnamefont{R.~U.} \bibnamefont{{Sexl}}},
  \bibinfo{journal}{\grg} \textbf{\bibinfo{volume}{8}}, \bibinfo{pages}{515}
  (\bibinfo{year}{1977}{\natexlab{b}}).

\bibitem[{\citenamefont{{Mansouri} and
  {Sexl}}(1977{\natexlab{c}})}]{mansouri77c}
\bibinfo{author}{\bibfnamefont{R.}~\bibnamefont{{Mansouri}}} \bibnamefont{and}
  \bibinfo{author}{\bibfnamefont{R.~U.} \bibnamefont{{Sexl}}},
  \bibinfo{journal}{\grg} \textbf{\bibinfo{volume}{8}}, \bibinfo{pages}{809}
  (\bibinfo{year}{1977}{\natexlab{c}}).

\bibitem[{\citenamefont{{Vargas} and {Torr}}(1989)}]{vargas89}
\bibinfo{author}{\bibfnamefont{J.~G.} \bibnamefont{{Vargas}}} \bibnamefont{and}
  \bibinfo{author}{\bibfnamefont{D.~G.} \bibnamefont{{Torr}}},
  \bibinfo{journal}{\pra} \textbf{\bibinfo{volume}{39}}, \bibinfo{pages}{2878}
  (\bibinfo{year}{1989}).

\bibitem[{\citenamefont{{Will}}(1992)}]{will92}
\bibinfo{author}{\bibfnamefont{C.~M.} \bibnamefont{{Will}}},
  \bibinfo{journal}{\prd} \textbf{\bibinfo{volume}{45}}, \bibinfo{pages}{403}
  (\bibinfo{year}{1992}).

\bibitem[{\citenamefont{Vetharaniam and Stedman}(1993)}]{vetharaniam93}
\bibinfo{author}{\bibfnamefont{I.}~\bibnamefont{Vetharaniam}} \bibnamefont{and}
  \bibinfo{author}{\bibfnamefont{G.~E.} \bibnamefont{Stedman}},
  \bibinfo{journal}{\pl} \textbf{\bibinfo{volume}{A~183}}, \bibinfo{pages}{349}
  (\bibinfo{year}{1993}).

\bibitem[{\citenamefont{Einstein}(1905)}]{einstein05}
\bibinfo{author}{\bibfnamefont{A.}~\bibnamefont{Einstein}},
  \bibinfo{journal}{\annphys} \textbf{\bibinfo{volume}{17}},
  \bibinfo{pages}{891} (\bibinfo{year}{1905}).

\bibitem[{\citenamefont{Minguzzi}(2002)}]{minguzzi02}
\bibinfo{author}{\bibfnamefont{E.}~\bibnamefont{Minguzzi}},
  \bibinfo{journal}{\fpl} \textbf{\bibinfo{volume}{15}}, \bibinfo{pages}{153}
  (\bibinfo{year}{2002}).

\bibitem[{\citenamefont{{Minguzzi}}(2003)}]{minguzzi03}
\bibinfo{author}{\bibfnamefont{E.}~\bibnamefont{{Minguzzi}}},
  \bibinfo{journal}{\cqg} \textbf{\bibinfo{volume}{20}}, \bibinfo{pages}{2443}
  (\bibinfo{year}{2003}).

\bibitem[{\citenamefont{Goy}(1996{\natexlab{a}})}]{goy96b}
\bibinfo{author}{\bibfnamefont{F.}~\bibnamefont{Goy}}, \bibinfo{journal}{\grqc}
   (\bibinfo{year}{1996}{\natexlab{a}}), \bibinfo{note}{{London conference on
  Physical Interpretations of Relativity Theory}}, \eprint{gr-qc/9607047}.

\bibitem[{\citenamefont{Goy}(1996{\natexlab{b}})}]{goy96c}
\bibinfo{author}{\bibfnamefont{F.}~\bibnamefont{Goy}}, \bibinfo{journal}{\grqc}
   (\bibinfo{year}{1996}{\natexlab{b}}), \eprint{gr-qc/9607042}.

\bibitem[{\citenamefont{Goy}(1997{\natexlab{a}})}]{goy97}
\bibinfo{author}{\bibfnamefont{F.}~\bibnamefont{Goy}}, \bibinfo{journal}{\grqc}
   (\bibinfo{year}{1997}{\natexlab{a}}), \eprint{gr-qc/9702042}.

\bibitem[{\citenamefont{Goy}(1997{\natexlab{b}})}]{goy97d}
\bibinfo{author}{\bibfnamefont{F.}~\bibnamefont{Goy}}, \bibinfo{journal}{\grqc}
   (\bibinfo{year}{1997}{\natexlab{b}}), \bibinfo{note}{{Talk of the Athen's
  Conference on Relativistic Physics}}, \eprint{gr-qc/9707005}.

\bibitem[{\citenamefont{{Spavieri}}(1986)}]{spavieri86}
\bibinfo{author}{\bibfnamefont{G.}~\bibnamefont{{Spavieri}}},
  \bibinfo{journal}{\pra} \textbf{\bibinfo{volume}{34}}, \bibinfo{pages}{1708}
  (\bibinfo{year}{1986}).

\bibitem[{\citenamefont{{Croca} and {Selleri}}(1999)}]{croca99}
\bibinfo{author}{\bibfnamefont{J.~R.} \bibnamefont{{Croca}}} \bibnamefont{and}
  \bibinfo{author}{\bibfnamefont{F.}~\bibnamefont{{Selleri}}},
  \bibinfo{journal}{\ncim} \textbf{\bibinfo{volume}{B~114}},
  \bibinfo{pages}{447} (\bibinfo{year}{1999}).

\bibitem[{\citenamefont{{Wang} et~al.}(2003)\citenamefont{{Wang}, {Zheng},
  {Yao}, and {Langley}}}]{wang03}
\bibinfo{author}{\bibfnamefont{R.}~\bibnamefont{{Wang}}},
  \bibinfo{author}{\bibfnamefont{Y.}~\bibnamefont{{Zheng}}},
  \bibinfo{author}{\bibfnamefont{A.}~\bibnamefont{{Yao}}}, \bibnamefont{and}
  \bibinfo{author}{\bibfnamefont{D.}~\bibnamefont{{Langley}}},
  \bibinfo{journal}{\pl} \textbf{\bibinfo{volume}{A~312}}, \bibinfo{pages}{7}
  (\bibinfo{year}{2003}).

\bibitem[{\citenamefont{{Tartaglia} and {Ruggiero}}(2004)}]{tartaglia04}
\bibinfo{author}{\bibfnamefont{A.}~\bibnamefont{{Tartaglia}}} \bibnamefont{and}
  \bibinfo{author}{\bibfnamefont{M.~L.} \bibnamefont{{Ruggiero}}},
  \bibinfo{journal}{\grqc}  (\bibinfo{year}{2004}), \eprint{gr-qc/0401005}.

\bibitem[{\citenamefont{{Dingle}}(1956)}]{dingle56}
\bibinfo{author}{\bibfnamefont{H.}~\bibnamefont{{Dingle}}},
  \bibinfo{journal}{\ppsa} \textbf{\bibinfo{volume}{69}}, \bibinfo{pages}{925}
  (\bibinfo{year}{1956}).

\bibitem[{\citenamefont{{Brans} and {Stewart}}(1973)}]{brans73}
\bibinfo{author}{\bibfnamefont{C.~H.} \bibnamefont{{Brans}}} \bibnamefont{and}
  \bibinfo{author}{\bibfnamefont{D.~R.} \bibnamefont{{Stewart}}},
  \bibinfo{journal}{\prd} \textbf{\bibinfo{volume}{8}}, \bibinfo{pages}{1662}
  (\bibinfo{year}{1973}).

\bibitem[{\citenamefont{{Low}}(1990)}]{low90}
\bibinfo{author}{\bibfnamefont{R.~J.} \bibnamefont{{Low}}},
  \bibinfo{journal}{\ejp} \textbf{\bibinfo{volume}{11}}, \bibinfo{pages}{25}
  (\bibinfo{year}{1990}).

\bibitem[{\citenamefont{{Dray}}(1990)}]{dray90}
\bibinfo{author}{\bibfnamefont{T.}~\bibnamefont{{Dray}}},
  \bibinfo{journal}{\ajp} \textbf{\bibinfo{volume}{58}}, \bibinfo{pages}{822}
  (\bibinfo{year}{1990}).

\bibitem[{\citenamefont{{Peters}}(1983)}]{peters83}
\bibinfo{author}{\bibfnamefont{P.~C.} \bibnamefont{{Peters}}},
  \bibinfo{journal}{\ajp} \textbf{\bibinfo{volume}{51}}, \bibinfo{pages}{791}
  (\bibinfo{year}{1983}).

\bibitem[{\citenamefont{{Barrow} and {Levin}}(2001)}]{barrow01}
\bibinfo{author}{\bibfnamefont{J.~D.} \bibnamefont{{Barrow}}} \bibnamefont{and}
  \bibinfo{author}{\bibfnamefont{J.}~\bibnamefont{{Levin}}},
  \bibinfo{journal}{\pra} \textbf{\bibinfo{volume}{63}},
  \bibinfo{pages}{044104} (\bibinfo{year}{2001}).

\bibitem[{\citenamefont{{Uzan} et~al.}(2002)\citenamefont{{Uzan}, {Luminet},
  {Lehoucq}, and {Peter}}}]{uzan02}
\bibinfo{author}{\bibfnamefont{J.}~\bibnamefont{{Uzan}}},
  \bibinfo{author}{\bibfnamefont{J.}~\bibnamefont{{Luminet}}},
  \bibinfo{author}{\bibfnamefont{R.}~\bibnamefont{{Lehoucq}}},
  \bibnamefont{and} \bibinfo{author}{\bibfnamefont{P.}~\bibnamefont{{Peter}}},
  \bibinfo{journal}{\ejp} \textbf{\bibinfo{volume}{23}}, \bibinfo{pages}{277}
  (\bibinfo{year}{2002}).

\bibitem[{\citenamefont{Ehrenfest}(1909)}]{ehrenfest09}
\bibinfo{author}{\bibfnamefont{P.}~\bibnamefont{Ehrenfest}},
  \bibinfo{journal}{\pz} \textbf{\bibinfo{volume}{10}}, \bibinfo{pages}{918}
  (\bibinfo{year}{1909}).

\bibitem[{\citenamefont{Rizzi and Ruggiero}(2002)}]{rizzi02}
\bibinfo{author}{\bibfnamefont{G.}~\bibnamefont{Rizzi}} \bibnamefont{and}
  \bibinfo{author}{\bibfnamefont{M.~L.} \bibnamefont{Ruggiero}},
  \bibinfo{journal}{\fpl} \textbf{\bibinfo{volume}{32}}, \bibinfo{pages}{1525}
  (\bibinfo{year}{2002}), \eprint{gr-qc/0207104}.

\bibitem[{\citenamefont{Gr{\o}n}(2004)}]{groen04}
\bibinfo{author}{\bibfnamefont{{\O}.}~\bibnamefont{Gr{\o}n}},
  \emph{\bibinfo{title}{Space geometry in rotating references frames: A
  historical appraisal}}, in  \cite{rrf04} (\bibinfo{year}{2004}).

\bibitem[{\citenamefont{Einstein}(1917)}]{einstein17}
\bibinfo{author}{\bibfnamefont{A.}~\bibnamefont{Einstein}},
  \emph{\bibinfo{title}{{\"Uber die spezielle und die allgemeine
  Relativit\"atstheorie}}} (\bibinfo{publisher}{Vieweg, Braunschweig},
  \bibinfo{year}{1917}), \bibinfo{note}{english in \ocite{einstein31}}.

\bibitem[{\citenamefont{Einstein}(1931)}]{einstein31}
\bibinfo{author}{\bibfnamefont{A.}~\bibnamefont{Einstein}},
  \emph{\bibinfo{title}{{Relativity; the special and general theory}}}
  (\bibinfo{publisher}{P. Smith, New York}, \bibinfo{year}{1931}),
  \bibinfo{note}{english version of \ocite{einstein17}}.

\bibitem[{\citenamefont{Einstein}(1921)}]{einstein21}
\bibinfo{author}{\bibfnamefont{A.}~\bibnamefont{Einstein}},
  \emph{\bibinfo{title}{{The Meaning of Relativity --- A collection of four
  lectures delivered at Princeton University}}} (\bibinfo{publisher}{Princeton
  University Press}, \bibinfo{year}{1921}), \bibinfo{note}{german version in
  \ocite{einstein22}}.

\bibitem[{\citenamefont{Einstein}(1922)}]{einstein22}
\bibinfo{author}{\bibfnamefont{A.}~\bibnamefont{Einstein}},
  \emph{\bibinfo{title}{{Vier Vorlesungen \"uber Relativit\"atstheorie,
  gehalten im Mai 1921 an der Universit\"at Princeton}}}
  (\bibinfo{publisher}{Vieweg, Braunschweig}, \bibinfo{year}{1922}),
  \bibinfo{note}{german version of \ocite{einstein21}}.

\bibitem[{\citenamefont{Einstein}(1956)}]{einstein56}
\bibinfo{author}{\bibfnamefont{A.}~\bibnamefont{Einstein}},
  \emph{\bibinfo{title}{{Grundz\"uge der Relativit\"atstheorie}}}
  (\bibinfo{publisher}{Vieweg, Braunschweig}, \bibinfo{year}{1956}),
  \bibinfo{note}{3. edition of \ocite{einstein22}}.

\bibitem[{\citenamefont{Einstein}(1911)}]{einstein11}
\bibinfo{author}{\bibfnamefont{A.}~\bibnamefont{Einstein}},
  \bibinfo{journal}{\pz} \textbf{\bibinfo{volume}{12}}, \bibinfo{pages}{509}
  (\bibinfo{year}{1911}).

\bibitem[{\citenamefont{{Peres}}(2004)}]{peres04}
\bibinfo{author}{\bibfnamefont{A.}~\bibnamefont{{Peres}}},
  \bibinfo{journal}{\grqc}  (\bibinfo{year}{2004}), \bibinfo{note}{submitted to
  \ajp}, \eprint{gr-qc/0401043}.

\bibitem[{\citenamefont{Ruggiero}(2003)}]{ruggiero03}
\bibinfo{author}{\bibfnamefont{M.~L.} \bibnamefont{Ruggiero}},
  \bibinfo{journal}{\ejp} \textbf{\bibinfo{volume}{24}}, \bibinfo{pages}{563}
  (\bibinfo{year}{2003}).

\bibitem[{\citenamefont{Bell}(1976)}]{bell76}
\bibinfo{author}{\bibfnamefont{J.~S.} \bibnamefont{Bell}},
  \bibinfo{journal}{Progress in Scientific Culture}
  \textbf{\bibinfo{volume}{1}} (\bibinfo{year}{1976}), \bibinfo{note}{also in
  \ocite{bell87}, pp.\ 67--80}.

\bibitem[{\citenamefont{Dewan and Beran}(1959)}]{dewan59}
\bibinfo{author}{\bibfnamefont{E.}~\bibnamefont{Dewan}} \bibnamefont{and}
  \bibinfo{author}{\bibfnamefont{M.}~\bibnamefont{Beran}},
  \bibinfo{journal}{\ajp} \textbf{\bibinfo{volume}{27}}, \bibinfo{pages}{517}
  (\bibinfo{year}{1959}).

\bibitem[{\citenamefont{Bell}(1987)}]{bell87}
\bibinfo{author}{\bibfnamefont{J.~S.} \bibnamefont{Bell}},
  \emph{\bibinfo{title}{Speakable and unspeakable in quantum mechanics}}
  (\bibinfo{publisher}{Cambridge University Press}, \bibinfo{year}{1987}).

\bibitem[{\citenamefont{{Field}}(2004)}]{field04}
\bibinfo{author}{\bibfnamefont{J.~H.} \bibnamefont{{Field}}},
  \bibinfo{journal}{\physics}  (\bibinfo{year}{2004}),
  \eprint{physics/0403094}.

\bibitem[{\citenamefont{Klauber}(1998)}]{klauber98}
\bibinfo{author}{\bibfnamefont{R.~D.} \bibnamefont{Klauber}},
  \bibinfo{journal}{\fpl} \textbf{\bibinfo{volume}{11}}, \bibinfo{pages}{405}
  (\bibinfo{year}{1998}), \eprint{gr-qc/0103076}.

\bibitem[{\citenamefont{Tartaglia}(1999)}]{tartaglia99}
\bibinfo{author}{\bibfnamefont{A.}~\bibnamefont{Tartaglia}},
  \bibinfo{journal}{\fpl} \textbf{\bibinfo{volume}{12}}, \bibinfo{pages}{17}
  (\bibinfo{year}{1999}), \eprint{physics/9808001}.

\bibitem[{\citenamefont{Tartaglia}(2004)}]{tartaglia04b}
\bibinfo{author}{\bibfnamefont{A.}~\bibnamefont{Tartaglia}},
  \emph{\bibinfo{title}{Does anything happen on a rotating disk?}}, in
  \cite{rrf04} (\bibinfo{year}{2004}).

\bibitem[{\citenamefont{Ashby}(2004)}]{ashby04}
\bibinfo{author}{\bibfnamefont{N.}~\bibnamefont{Ashby}},
  \emph{\bibinfo{title}{The Sagnac effect in the Global Positioning System}},
  in  \cite{rrf04} (\bibinfo{year}{2004}).

\bibitem[{\citenamefont{Minguzzi and Macdonald}(2003)}]{minguzzi03b}
\bibinfo{author}{\bibfnamefont{E.}~\bibnamefont{Minguzzi}} \bibnamefont{and}
  \bibinfo{author}{\bibfnamefont{A.}~\bibnamefont{Macdonald}},
  \bibinfo{journal}{\fpl} \textbf{\bibinfo{volume}{16}}, \bibinfo{pages}{593}
  (\bibinfo{year}{2003}).

\bibitem[{\citenamefont{Selleri}(1996)}]{selleri96}
\bibinfo{author}{\bibfnamefont{F.}~\bibnamefont{Selleri}},
  \bibinfo{journal}{\fp} \textbf{\bibinfo{volume}{26}}, \bibinfo{pages}{641}
  (\bibinfo{year}{1996}).

\bibitem[{\citenamefont{Selleri}(1997)}]{selleri97}
\bibinfo{author}{\bibfnamefont{F.}~\bibnamefont{Selleri}},
  \bibinfo{journal}{\fpl} \textbf{\bibinfo{volume}{10}}, \bibinfo{pages}{72}
  (\bibinfo{year}{1997}).

\bibitem[{\citenamefont{Rizzi and Serafini}(2004)}]{rizzi04b}
\bibinfo{author}{\bibfnamefont{G.}~\bibnamefont{Rizzi}} \bibnamefont{and}
  \bibinfo{author}{\bibfnamefont{A.}~\bibnamefont{Serafini}},
  \emph{\bibinfo{title}{Synchronization and Desynchronization on Rotating
  Platforms}}, in  \cite{rrf04} (\bibinfo{year}{2004}).

\bibitem[{\citenamefont{Weber}(2004)}]{weber04}
\bibinfo{author}{\bibfnamefont{T.~A.} \bibnamefont{Weber}},
  \emph{\bibinfo{title}{Elementary Considerations of the Time and Geometry of
  Rotating Reference Frames}}, in  \cite{rrf04} (\bibinfo{year}{2004}).

\bibitem[{\citenamefont{Bergia and Guidone}(1998)}]{bergia98}
\bibinfo{author}{\bibfnamefont{S.}~\bibnamefont{Bergia}} \bibnamefont{and}
  \bibinfo{author}{\bibfnamefont{M.}~\bibnamefont{Guidone}},
  \bibinfo{journal}{\fpl} \textbf{\bibinfo{volume}{11}}, \bibinfo{pages}{549}
  (\bibinfo{year}{1998}).

\bibitem[{\citenamefont{Verne}(1873)}]{verne1873}
\bibinfo{author}{\bibfnamefont{J.}~\bibnamefont{Verne}},
  \emph{\bibinfo{title}{{Tour du monde en quatre-vingts jours}}}
  (\bibinfo{publisher}{Pierre-Jules Hetzel, Paris}, \bibinfo{year}{1873}),
  \bibinfo{note}{{English: \emph{Around the world in 80 days}}}.

\bibitem[{\citenamefont{Goy and Selleri}(1997)}]{goy97c}
\bibinfo{author}{\bibfnamefont{F.}~\bibnamefont{Goy}} \bibnamefont{and}
  \bibinfo{author}{\bibfnamefont{F.}~\bibnamefont{Selleri}},
  \bibinfo{journal}{\fpl} \textbf{\bibinfo{volume}{10}}, \bibinfo{pages}{17}
  (\bibinfo{year}{1997}), \eprint{gr-qc/9702055}.

\bibitem[{\citenamefont{Klauber}(2003)}]{klauber03}
\bibinfo{author}{\bibfnamefont{R.~D.} \bibnamefont{Klauber}},
  \bibinfo{journal}{\fpl} \textbf{\bibinfo{volume}{16}}, \bibinfo{pages}{447}
  (\bibinfo{year}{2003}).

\bibitem[{\citenamefont{Selleri}(2004)}]{selleri04}
\bibinfo{author}{\bibfnamefont{F.}~\bibnamefont{Selleri}},
  \emph{\bibinfo{title}{Sagnac effect: end of the mystery}}, in  \cite{rrf04}
  (\bibinfo{year}{2004}).

\bibitem[{\citenamefont{Rizzi and Tartaglia}(1998)}]{rizzi98}
\bibinfo{author}{\bibfnamefont{G.}~\bibnamefont{Rizzi}} \bibnamefont{and}
  \bibinfo{author}{\bibfnamefont{A.}~\bibnamefont{Tartaglia}},
  \bibinfo{journal}{\fp} \textbf{\bibinfo{volume}{28}}, \bibinfo{pages}{1663}
  (\bibinfo{year}{1998}).

\bibitem[{\citenamefont{Rizzi and Tartaglia}(1999)}]{rizzi99}
\bibinfo{author}{\bibfnamefont{G.}~\bibnamefont{Rizzi}} \bibnamefont{and}
  \bibinfo{author}{\bibfnamefont{A.}~\bibnamefont{Tartaglia}},
  \bibinfo{journal}{\fpl} \textbf{\bibinfo{volume}{12}}, \bibinfo{pages}{179}
  (\bibinfo{year}{1999}).

\bibitem[{\citenamefont{{Barrow} and {Levin}}(2003)}]{barrow03}
\bibinfo{author}{\bibfnamefont{J.~D.} \bibnamefont{{Barrow}}} \bibnamefont{and}
  \bibinfo{author}{\bibfnamefont{J.}~\bibnamefont{{Levin}}},
  \bibinfo{journal}{\mnras} \textbf{\bibinfo{volume}{346}},
  \bibinfo{pages}{615} (\bibinfo{year}{2003}).

\bibitem[{\citenamefont{{Anandan}}(1981)}]{anandan81}
\bibinfo{author}{\bibfnamefont{J.}~\bibnamefont{{Anandan}}},
  \bibinfo{journal}{\prd} \textbf{\bibinfo{volume}{24}}, \bibinfo{pages}{338}
  (\bibinfo{year}{1981}).

\bibitem[{\citenamefont{Anandan and Suzuki}(2004)}]{anandan04}
\bibinfo{author}{\bibfnamefont{J.}~\bibnamefont{Anandan}} \bibnamefont{and}
  \bibinfo{author}{\bibfnamefont{J.}~\bibnamefont{Suzuki}},
  \emph{\bibinfo{title}{Quantum Mechanics in a Rotating Frame}}, in
  \cite{rrf04} (\bibinfo{year}{2004}).

\bibitem[{\citenamefont{Davidovi{\'c} and Arsenovi{\'c}}(2004)}]{davidovic04}
\bibinfo{author}{\bibfnamefont{D.~M.} \bibnamefont{Davidovi{\'c}}}
  \bibnamefont{and}
  \bibinfo{author}{\bibfnamefont{D.}~\bibnamefont{Arsenovi{\'c}}},
  \bibinfo{journal}{\fpl} \textbf{\bibinfo{volume}{17}}, \bibinfo{pages}{183}
  (\bibinfo{year}{2004}).

\end{thebibliography}

\end{document}